\RequirePackage[hyphens]{url}
\documentclass[12pt, oneside]{amsart}
\pdfoutput=1
\usepackage{geometry}
\usepackage{lmodern} 
\usepackage{amsmath}
\usepackage{amssymb}
\usepackage{amsthm}
\usepackage{comment}
\usepackage{color}
\usepackage[usenames,dvipsnames]{xcolor}
\usepackage{graphicx}
\usepackage{tikz}
\usetikzlibrary{shapes,snakes}
\usepackage{subfig} 
\usepackage{soul}
\usepackage[countmax]{subfloat}
\usepackage{float}
\usepackage{rotfloat}
\usepackage{setspace}
\usepackage{esint}
\usepackage[authoryear]{natbib}
\definecolor{MyDarkBlue}{rgb}{0,0.08,0.45}
\usepackage[unicode=true,pdfusetitle, bookmarks=true,bookmarksnumbered=false,bookmarksopen=false, breaklinks=false,pdfborder={0 0 1},backref=section,colorlinks=true, linkcolor = MyDarkBlue, urlcolor = MyDarkBlue, citecolor = MyDarkBlue]{hyperref}
\usepackage[para]{threeparttable}
\usepackage{verbatim}
\usepackage{fancyvrb}
\usepackage{pdflscape}

\PassOptionsToPackage{normalem}{ulem}
\usepackage{ulem}

\makeatletter

\numberwithin{equation}{section}

\hypersetup{colorlinks=true, linkcolor=MyDarkBlue}

\newtheorem{rem}{{\bf \sc Remark}}

\providecommand{\E}{\mathrm{E}}

\providecommand{\Prob}{\mathrm{P}}

\renewcommand{\Pr}{\Prob}

\makeatother

\begin{document}

\title[When Celebrities Speak]{When Celebrities Speak:\\ A Nationwide Twitter Experiment Promoting Vaccination in Indonesia}

\author{Vivi Alatas$^{\dagger}$ }
\author{Arun G. Chandrasekhar$^{\ddagger}$}
\author{Markus Mobius$^{\S}$ }
\author{Benjamin A. Olken$^{\star}$ }
\author{Cindy Paladines$^{\dagger}$ }
\date{\today}

\thanks{We thank Marcella Alsan, Nancy Baym, Emily Breza, Leo Bursztyn, Rebecca Diamond, Dean Eckles, Paul Goldsmith-Pinkham, Ben Golub, Rema Hanna, Mary Gray, Matt Jackson, Matthew Wai-Poi, Alex Wolitsky, and participants at various seminars for helpful discussions. Aaron Kaye, Nurzanty Khadijah, Devika Lakhote, Eva Lyubich, Sinead Maguire, Lina Marliani, Sebastian Steffen, Vincent Tanutama provided excellent research assistance. We thank Nila Moeloek, then Indonesian Special Envoy for Sustainable Development Goals, Diah Saminarsih, and their team for providing support for this project. This study was approved by IRBs at MIT (Protocol \#1406006433) and Stanford (Protocol \#31451), and registered in the AEA Social Science Registry (AEARCTR-0000757).  Funding for this project came from the Australian Government Department of Foreign Affairs and Trade. 
}
\thanks{$^{\dagger}$World Bank}
\thanks{$^{\ddagger}$Department of Economics, Stanford University; NBER; J-PAL}
\thanks{$^{\S}$Microsoft Research, New England}
\thanks{$^{\star}$Department of Economics, MIT; NBER; J-PAL}

\maketitle

\begin{abstract}
Celebrity endorsements are often sought to influence public opinion. We ask whether celebrity endorsement per se has an effect beyond the fact that their statements are seen by many, and whether on net their statements actually lead people to change their beliefs. To do so, we conducted a nationwide Twitter experiment in Indonesia with 46 high-profile celebrities and organizations, with a total of 7.8 million followers, who agreed to let us randomly tweet or retweet content promoting immunization from their accounts. Our design exploits the structure of what information is passed on along a retweet chain on Twitter to parse reach versus endorsement effects. Endorsements matter: tweets that users can identify as being originated by a celebrity are far more likely to be liked or retweeted by users than similar tweets seen by the same users but without the celebrities' imprimatur. By contrast, explicitly citing sources in the tweets actually reduces diffusion. By randomizing which celebrities tweeted when, we find suggestive evidence that overall exposure to the campaign may influence beliefs about vaccination and knowledge of immunization-seeking behavior by one's network. Taken together, the findings suggest an important role for celebrity endorsement.


\end{abstract}

\setcounter{page}{0}

\newpage

\section{Introduction}
Social media has allowed celebrities to take an increasing role in social discourse. With millions of online followers, celebrities have a direct channel to spread messages on a wide variety of issues, many of which are far removed from their original reason for fame. Their very participation in ongoing discussions can make issues prominent and shape the zeitgeist.

Examples abound. \#BlackLivesMatter, a campaign against racial injustice, is the most-used social issue Twitter hashtag of all time, with 41 million uses as of September 2018. 
Prominent celebrity users include, among others, LeBron James, Drake, Carmelo Anthony, Kim Kardashian, Colin Kaepernick, Lin-Manuel Miranda, Kerry Washington, Kanye West, Serena Williams, and Zendaya (with over 211 million non-unique Twitter followers among them).   In public health, the \#IceBucketChallenge, promoting awareness of Lou Gehrig's disease, became the sixth most used social issue hashtag of all time following participation by a wide range of celebrities, from Oprah to Bill Gates. Each of these campaigns was initiated by a less-well-known activist, but was made prominent in part through celebrity participation.\footnote{Campaigns are typically initiated by lesser-known activists. \#BlackLivesMatter was created by writers and activists Alicia Garza, Patrisse Cullors, and Opal Tomet  and \#IceBucketChallenge by Peter Frates.} As a result, policymakers and firms often seek out celebrity endorsements, whether to advance public-interest causes or to promote products.

Key questions, however,  are both \emph{whether} and \emph{why} certain campaigns are so effective. The first issue concerns why celebrities matter so much. Celebrities have broad \emph{reach}---people are watching what Kim Kardashian says or does, and hence her actions and utterances are seen by many people. Moreover, celebrities may have an \emph{endorsement} power above and beyond their reach. The fact that Kim Kardashian was willing to endorse a given product or cause may lead people to update about the quality of the product or the importance of that cause, or people may simply want to be like her.\footnote{Of course, it is  possible that the endorsement effect could be negative.} If the endorsement effect is present, this means that celebrities have an outsized importance: it is not just that they reach so many people, but their voice \textit{per se} has an additional effect. 

More generally, the messaging itself may vary in credibility in ways beyond whether it is explicitly endorsed. For instance, if the tweets used systematically cited verifiable sources, the information would be less subject to doubt and presumably carry more value. On the other hand, it is possible that inclusion of credible sources itself discourages passing on the message. It is also possible that hearing information from multiple different sources may be more powerful than hearing it from one person. If so, that implies an additional role for celebrities, as seeding information with very central individuals such as celebrities makes it more likely that people will hear information from multiple sources. 

Understanding the nature of social influence is challenging for several reasons. First and foremost, celebrities' decisions about whether to make public statements are endogenously determined and influenced by the general information environment into which they are speaking. People also consume information from such a wide variety of sources that is also near impossible to isolate the impact of information from a particular source on overall beliefs. Even if one could credibly solve the endogeneity problem of whether celebrities choose to speak on a topic, and could isolate the impact on a particular individual, a given action by a celebrity bundles reach and endorsement effects, making it hard to disentangle why, precisely, these messages have an impact. Moreover, the precise choice of content and whether the content involves citing or linking to a source all are subject to endogeneity concerns.

To study these issues, we conducted an experiment through a nationwide immunization campaign on Twitter from 2015-2016 in Indonesia, in collaboration with the Indonesian Government's Special Ambassador to the United Nations for Millennium Development Goals. Working with the Special Ambassador, we recruited 46 high-profile celebrities and organizations, with a total of over 7.8 million followers, each of whom gave us access to send up to 33 tweets or retweets promoting immunization from their accounts. The content and timing of each of these tweets was randomly chosen by us from a set of tweets approved by the Indonesian Ministry of Health, all of which featured a campaign hashtag \#AyoImunisasi (``Let's Immunize''). All our participants joined knowing that they would not be able to affect the text or timing of the tweets.\footnote{Celebrities were allowed to a veto a tweet if they did not want it sent from their account, though this in fact never happened during the campaign.} 

The experiment randomly varied the tweets along three dimensions: (1) Did the celebrity / organization send the tweet from their account, or did they retweet a message (drawn randomly from the same tweet library) sent by us from an an ordinary (non-celebrity) user's account?; (2) Did the tweet explicitly cite a source to bolster its credibility?; (3) Which days of the campaign did this influencer-tweet/retweet event happen?

The random variation allows us to causally address two sets of  questions. First, can we understand whether and why celebrity-involved campaigns have influence? Is it because of the reach of the celebrity, the endorsement effect, the content of the message, or the extent of exposure that they induce through their messaging?  Second, does exposure to an online information campaign about a public health topic lead to changes in beliefs and offline behavior, through generating conversations and perhaps health-seeking responses?

We chose this setting for several reasons. Indonesia is very active on social media; for example, in 2012 its capital, Jakarta, originated the most tweets of any city in the world. Twitter also has a number of useful features for our study. Because both the network (i.e., who sees whose tweets) and virtually all information flows over the network (i.e., tweets and retweets) on Twitter are public information, we can precisely map which individual sees what information, as well as where they saw it from, allowing us to observe for each user how much exposure they had to precise bits of information. By conducting an experiment, in which we randomly vary who tweets what when, we can both solve the identification problem of endogenous speaking behavior, as well as disentangle reach vs. endorsement effects. Finally, again because of the public nature of Twitter, we can observe people's responses to information both online (by observing their online ``liking'' and ``retweeting'' behavior) and offline (by conducting a phone survey of Twitter users and linking their survey responses to what they saw on Twitter.) 

Beyond being a useful lab for our study, Twitter is one of the most important mediums of information exchange in the world. With over 1 billion users and 328 million active users, Twitter provides a platform for individuals to broadcast messages widely. Celebrities, politicians, and organizations are widely followed.\footnote{Among the most followed worldwide are Katy Perry (107 million), Barack Obama (104 million), Ellen DeGeneres (77 million), Kim Kardashian (59 million), Donald Trump (58 million), CNN Breaking News (55 million),  Bill Gates (46 million), Narendra Modi (45 million), The New York Times (43 million), LeBron James (42 million),  and Shah Rukh Khan (37 million) as of January 2019.} 
As such, influencers have a platform to directly message en masse and engage on timely issues. 

We begin with our core question -- unpacking why and how celebrity messages may matter -- using variation we induced within celebrities in the nature and content of their tweets. To tease out the role of celebrity endorsement per se, our design exploits the unique structure of how information is passed on Twitter.  Messages in Twitter are passed on by retweeting a message to one's followers.  Crucially for our design, when a message is retweeted, the follower observes who originally composed the tweet, and who retweeted it directly to the follower, but not any intermediate steps in the path. 

We exploit this feature to distinguish reach from endorsement. Consider the difference between what happens when 1) we have a celebrity directly compose and tweet a message, compared to 2) when we have a celebrity retweet a message drawn from the same pool of tweets but originated by a normal citizen (whom we henceforth denote as a ``ordinary Joes and Janes''; these Joes/Janes are also participants on our campaigns). 
	
In the first case, some celebrity followers (whom we denote $F_1$) retweet it to their followers, whom we denote $F_2$. The followers-of-followers ($F_2$s) observe that the celebrity authored the message and that $F_1$ retweeted it. But in the second case, when the celebrity retweeted a Joe/Jane's message rather than composed it herself, the followers-of-followers of the celebrity ($F_2$s) observe only that the Joe/Jane tweeted and that $F_1$ then retweeted for $F_2$ to see. Notice that in this way, $F_2$ is randomly blinded to the celebrity's involvement in the latter case, as compared to the former: differences in $F_2$'s behavior therefore correspond to differences due to knowing that the celebrity was involved.\footnote{A challenge in the design is that the $F_1$ decision to retweet may be endogenous. We discuss this issue in detail in Section \ref{subsec:value_endorsement} below, and show that the results are largely similar in the subset of cases where $F_1$s were also study participants whom we randomly selected and forced to retweet exogenously, and hence the sample of exposed $F_2$s is identical.} In the period we study, the ordering of the Twitter feed was strictly chronological, so this design manipulates whether the $F_2$s know about the celebrities involvement without affecting how prominently the message appeared in the Twitter feed. 

We study the impacts of this induced variation using online reactions to the tweets, i.e., likes and retweets, so we can observe the reactions of every individual follower to every specific tweet.\footnote{Recall that a ``like'' corresponds to simply clicking a button to indicate that one likes the message (and the action is not pushed to one's followers),  while ``retweet'' subsequently passes on the tweet to all of one's followers. While it is certainly the case, \textit{a priori}, that individuals may retweet tweets that they even disagree with, adding commentary or simply ironically, ``liking'' the tweet directly conveys approval.} 

We find strong evidence that the celebrity's endorsement \emph{per se} matters. In particular, we find using this design that when an individual observes a given message through a retweet, and that message was randomized to be composed by a celebrity as compared to an ordinary individual, there is an 70 percent increase in the number of likes and retweets, compared to similar messages when the celebrity's involvement was masked. We find similar results even when restrict attention to those cases where $F_1s$ are participants in the experiment and we exogenously had them retweet the message, ensuring that whether the $F_2$ was exposed to the message in the first place was completely exogenous. 

We then look to the role of source citation, exploiting our second randomization of whether a given tweet is randomly assigned to have a verifiable source attached to its claim or not. We find, perhaps surprisingly, that messages are less likely to be passed on if they are randomly assigned a source. This is true regardless of whether tweet was composed by the celebrity themselves, or composed by an ordinary Joe/Jane and retweeted by a celebrity. The magnitudes are substantial: for instance, randomly attaching a source to a tweet that the celebrity retweets corresponds to a 50 percent 
decline in the subsequent retweet rate. One interpretation is the information is less novel if it is sourced; more generally, we discuss theoretically how increasing the reliability of information passed has ex ante ambiguous effects on the probability the information is passed.  

The final piece of our online analysis is to examine exposure effects: does hearing a message multiple times (from multiple different sources) have linear, concave, or convex effects on the probability of passing on the message? This is important because if an individual passes on messages after a single exposure (simple contagion) versus requiring many exposures (complex contagion), the diffusion processes wind up being very different.  In the latter case central individuals such as celebrities may matter more. 
We find evidence consistent with complex contagion, but with concave effects: while going from one to two messages increases the probability of retweeting two-fold, and going from one to three increases the probability by 2.5-fold, the effect flattens out after that.

Taken together, the findings suggest an important set of considerations in policy design of a social media campaign. Celebrity involvement is crucial, not only for their direct broadcast effect but their endorsement effects as well. In contrast, perhaps counterintuitively, sources can actually slow down a campaign. And at the margin, efforts should be placed not to repeated messaging but rather to wider messaging: a budget of messages should be spread out to maximize repeated exposure.

Given that people seem to pay attention to the campaign online -- liking and retweeting celebrity messages on immunization -- a natural next question is whether such a campaign has effects on real-world beliefs, knowledge, and  behavior. To study this, we used the timing of the tweets to randomly generate differences in exposure to our campaign. Specifically, we randomized the celebrities into two groups, with the first group assigned to tweet during July and August 2015 (Phase I) and the second group assigned to tweet from November 2015 - February 2016 (Phases II and III). We conducted a phone survey of a subset of followers of our celebrities in between these two groups of tweets. Since we know which of our celebrities each of  these followers followed at baseline, this randomization into two phases generates random variation in how many immunization-related tweets from our campaign each individual had potentially been exposed to as of the time of our survey. 

The evidence using this variation, while suggestive, indicates that exposure to celebrity endorsements does have measurable effects. We begin by showing that people did pay attention: a one standard deviation increase in exposure to the campaign due to our randomization, equivalent to about 15 tweets or retweets showing up on a user's Twitter feed over a period of about one month, corresponds to a 20 percent increase in the probability that the respondent in the phone survey knows about our hashtag, \#AyoImunisasi; an 11 percent increase in the probability they have heard about immunization through Twitter; and a 14 percent increase in the number of times they report having heard about immunization through Twitter. We then show that exposure to the campaign  
may have increased knowledge about immunization. We asked phone respondents a number of factual questions about immunization (e.g., whether the vaccine was domestically produced, an important public message for the Government as domestically produced vaccines are known to be halal and hence allowed under Muslim law), all of which were addressed in some of the campaign tweets. A one standard deviation increase in exposure to the campaign corresponds to a 12 percent increase in the probability that the respondent knows that vaccines are domestic (on a base of 58 percent in knowledge in the whole sample), though no increase in the three other dimensions of knowledge we examined.

We then turn to respondents' knowledge of immunization behavior in their neighbor, friend, and relative networks. In particular, we ask whether knowledge of immunization behavior of members of each of these networks increased, which is a soft measure of offline discussion about immunization in their respective networks.  
 Again we find effects of the celebrity pro-immunization campaign: a one standard deviation increase in exposure corresponds to a 23 percent increase in the probability of knowing about one's neighbors' recent immunization behavior. We find no increases in knowledge for friends and relatives. The idea that one would learn about immunization decisions of neighbors is consistent with immunization practices in Indonesia, which take place at \textit{posyandu} meetings, staffed by a midwife, that occur each month in each neighborhood (\textit{dusun} or \textit{RW}) of Indonesia \citep{olken2014should}.  

We look at changes in reported immunization decisions of respondents and those in their network. We find no effects on one's own immunization decisions, though our statistical power is such that we cannot rule out substantial effects. This is not surprising, because it is unlikely that any given member of our sample had a child exactly in the duration of our study. But casting a wider net,  we find consistent evidence that in each type of network---neighbors, relatives, and friends---of those exposed to the campaign, those exposed to the Twitter campaign were more likely to report that their network members actually immunized their children. In sum, while the estimates in each domain are suggestive, taken together we find consistent evidence that celebrity endorsements actually may affect a combination of offline knowledge about facts and the knowledge of health status and health-seeking behavior by one's neighbor, friend, and relative network members.

\subsection*{Related Literature}  This work relates to a literature on the diffusion of information for public policy \citep*{ryan1943diffusion,kremer2007illusion,conley2010learning,katonazs2011,banerjee2013diffusion,beaman2016can}. To our knowledge, our paper represents the largest randomized controlled trial of an online diffusion experiment, particularly one that involves major influencers. Moreover, while this literature has studied the flow of information over social networks, and how position in the network affects the flow of information, it has typically been silent on whether the identity of the individual who passes on the information matters per se.\footnote{An important exception is \cite{beaman2018diffusion} who look at how gender plays a role in information diffusion.} Indeed, this is because normally the identity of an individual and that individual's position in the network go hand-in-hand, so varying who is sending the information changes both of these simultaneously; our experimental design, by contrast, allows us to separate these two effects. 

Moreover, unlike these previous studies, which have been carried out in smaller scales, we note that our study represents exactly the kind of public awareness campaign that governments and large-scale policymakers are interested in, as represented by the Ministry of Health's and World Bank's interest in partnership. To our knowledge, something like this has never been studied experimentally before. 

There is also a literature on generating online cascades  \citep{leskovec2007dynamics,bakshy2011everyone}. This literature follows online diffusions through Twitter, Facebook, and other social media, and through observational studies looks at what drives and does not drive diffusion. Much of the literature concludes that under a wide range of assumptions, it is more worthwhile to seed a message through a bunch of ordinary citizens as compared to identifying and targeting any particular influencer. As the research notes, however, there is no causal evidence for the role of influencers here, and certainly no causal evidence to parse what aspects of celebrity involvement matters. Of course, in observational studies, what celebrities say, whether they cite sources, whether they endorse others' messages are all endogenous. Our experiment allows us to move past this. Further, by linking our online behavior to offline beliefs and behavior, we can take a step towards measuring, albeit in a limited and minimal way, policy impact.

Furthermore, the paper speaks to a literature that has looked at how media exposure affects behavior, with an emphasis on civic engagement. The literature demonstrates that exposure to mass media such as newspapers, television, and radios can contribute to health-seeking behavior, political positioning, voting behavior, and economic behavior more generally  \citep*[see, e.g., ][]{alsan2017tuskegee,gentzkow2004media,dellavigna2007fox,enikolopov2011media,gentzkow2011effect,dellavigna2014cross,martin2017bias,bursztyn2016tear,allcott2017social}. 
Our paper moves to \emph{social} media, studying the online and offline effects of a large-scale social network campaign.\footnote{In that sense, our paper is related to \cite*{enikolopov2016social}, who use variation in the overall spread of a social network in Russia due to connections with the network's founder, and look at the effects of greater vs. less participation in social media on political outcomes. By contrast, our paper generates random variation in exposure within the social network. It is also related to \cite*{gong2017tweeting}, who experimentally vary tweets in China on Sina Weibo about TV programs and measure the impact on TV viewership.}

Finally, there has also been recent theoretical work exploring how (and whether) policymakers should  identify ``central'' individuals in a network to generate  diffusion.\footnote{See \citet*{kempekt2003,kempekt2005,banerjee2013diffusion,kimetal2015,beaman2016can,gossip2016,akbarpourjust}.} Our results show that (i) people are much more likely to pass on information if it originally came from a well-connected source (e.g., a celebrity), (ii) individuals are more likely to pass it on if they hear if from multiple distinct sources, (iii) there are only a few celebrities, and (iv) that an online public health campaign generates only short-timed diffusions. The results therefore confirm the importance of seeding information with influential people.

\subsection*{Organization} The rest of the paper is organized as follows. Section \ref{sec:experiment} lays out the setting, recruitment, experimental design, and sample statistics. In Section \ref{sec:value_endorsement}, we present our main results. We study the mechanisms of the  online diffusion process  and explore a variety of endorsement effects.   Then Section \ref{sec:offline} asks whether the increased online chatter corresponds to offline changes in beliefs  and behavior. Section \ref{sec:conclusion} concludes.

\

\section{Experiment}\label{sec:experiment}
\subsection{Setting and Sample}
Our study took place in Indonesia in 2015 and 2016. Despite being a developing country, Indonesia is quite active on social media and an excellent place to study social media dynamics. Indonesia ranks fourth worldwide in the number of Facebook accounts, with 126 million\footnote{\url{https://www.statista.com/statistics/268136/top-15-countries-based-on-number-of-facebook-users/}} in 2017 (about half the population); it also ranks third in the number of Twitter accounts, with over 16.8 million (about 6.4 percent of the population).\footnote{\url{https://www.statista.com/statistics/490548/twitter-users-indonesia/}} These Twitter users are active as well: in 2012, a study that linked individual tweets to their cities of origin found that Indonesia's capital, Jakarta, was the top city producing tweets anywhere in the world, narrowly exceeding Tokyo.\footnote{\url{https://semiocast.com/en/publications/2012_07_30_Twitter_reaches_half_a_billion_accounts_140m_in_the_US}}

The focus of the experiment was on improving immunization. Immunization was chosen as it was a government priority, as Indonesia was trying to improve its immunization rates as part of its drive to achieve the Millennium Development Goals. A set of 550 tweets was developed in close coordination with the Ministry of Health that sought to improve information about immunization. The tweets included information about access to immunization (i.e., immunizations are free, available at government clinics, and so on); information about the importance of immunization (i.e., immunizations are crucial to combat child diseases); and information designed to combat common myths about immunization (i.e., vaccines are made domestically in Indonesia and are therefore halal). For each tweet, we also identified a source (either a specific link or an organization's Twitter handle). All tweets were approved by the Ministry of Health, and all included a common hashtag, \#AyoImunisasi (``Let's Immunize''). Each tweet was written in Indonesian, and two versions were prepared---one using formal Indonesian, and one using casual/street Indonesian, to match the written tweeting styles of the participants.

With help from the Indonesian Special Ambassador to the United Nations for Millennium Development Goals, we recruited 37 high-profile Twitter users, whom we denote ``celebrities,'' with a total of 7.8 million Twitter followers, to participate in our experiment. These ``celebrities'' come from a wide range of backgrounds, including pop music stars, TV personalities, actors and actresses, motivational speakers, government officials, and public intellectuals. They have a mean of 262,647 Twitter followers each, with several having more than one million followers. We also recruited 9 organizations involved in public advocacy and/or health issues in Indonesia with a mean of 132,300 followers each. 

In addition to the celebrities, we recruited 1032 ordinary citizens, whom we call ``ordinary Joes and Janes''. The role of the Joes/Janes will be to allow us to have essentially unimportant, everyday individuals compose tweets that are then retweeted by celebrities, which will be important for identification. These Joes/Janes consist primarily of university students at a variety of Indonesian universities. They are far more typical in their Twitter profiles, with a mean of 511 followers.  

Every participant (both celebrities and Joes/Janes) consented to signing up with our app that (1) lets us tweet content from their account (13, 23, or 33 times), (2) randomize the content of the tweets from a large list of 549  
immunization tweets approved by the Ministry of Health, and (3) has no scope for editing. Participants were given two choices: (1) the maximum number of tweets to authorize (13, 23, or 33), 
and (2) a writing style for the tweet (to better approximate their normal writing style), either formal or slang language.\footnote{The website showed sample tweets to demonstrate the style.
}

\subsection{Experimental Design}\label{sec:design}

Our experiment is designed to understand which aspects of social media campaigns are important for disseminating a message.  
The choices we have at our disposal are (a) the originator of the message (a Joe/Jane or a celebrity), (b) whether the message contains a credible source, (c) the content of the message, and (d) whether the campaign should emphasize repeated messaging to the same people or more widespread messaging. Ex ante it may seem obvious, for instance, that sources are better (after all the information is more credible) and celebrity involvement is better (after all, for a variety of reasons the information may be viewed as more credible). But thinking carefully about the retweeting process demonstrates that, in fact, the effect of each of these design options is actually theoretically ambiguous, and hence ultimately an empirical question. Appendix \ref{sec:Model} presents an application of a simple model by \cite*{chandrasekhar2018signaling} to demonstrate the ambiguity, though certainly other  
models can be used.

The experimental design consists of the randomizing the content and timing of tweets and retweets among our participants. The design has two main components. First, to investigate the role of celebrity endorsements, we randomized the content of tweets and whether they were tweeted or retweeted by celebrities. We describe each of these in turn. Second, to measure the overall impact of the Twitter campaign on offline beliefs, we randomized celebrities into phases (i.e., in which months particular celebrities tweeted); this allows us to generate random variation in the amount of exposure a particular follower had received at the time of our survey.

First, within each phase, we randomized virtually all aspects of the tweets. Specifically, we randomized the precise timing of tweets (which day and what time of day); which tweet from our pre-prepared bank of approved tweets was tweeted by whom and when; whether a tweet was tweeted directly by a celebrity, or tweeted by a Joe/Jane and then retweeted by a celebrity; and, for a subset of tweets, whether the tweet included the `credibility boost' (i.e., the source link or referring organization's Twitter handle).\footnote{Note that in the period we study, a Twitter user saw all tweets and retweets from the users they follow in strict reverse chronological order (i.e., newest tweets appeared first, and so on). Twitter subsequently (in March 2016) applied an algorithm to prioritize the ordering of the tweets, but since in the period we study (July 2015 through February 2016) tweets appeared in strictly chronological order, our experimental design does not affect the ordering of tweets in a user's Twitter feed.} In addition, in Phase III, all tweets / retweets by a celebrity were then retweeted by a randomly selected number of Joes/Janes. These various randomizations allow us to identify the role of celebrity reach vs. endorsement, as well as the role of repeated exposure to information, as described in more detail in Section \ref{sec:value_endorsement} below. A schematic of the design is provided in Figure \ref{fig:expt_design}.

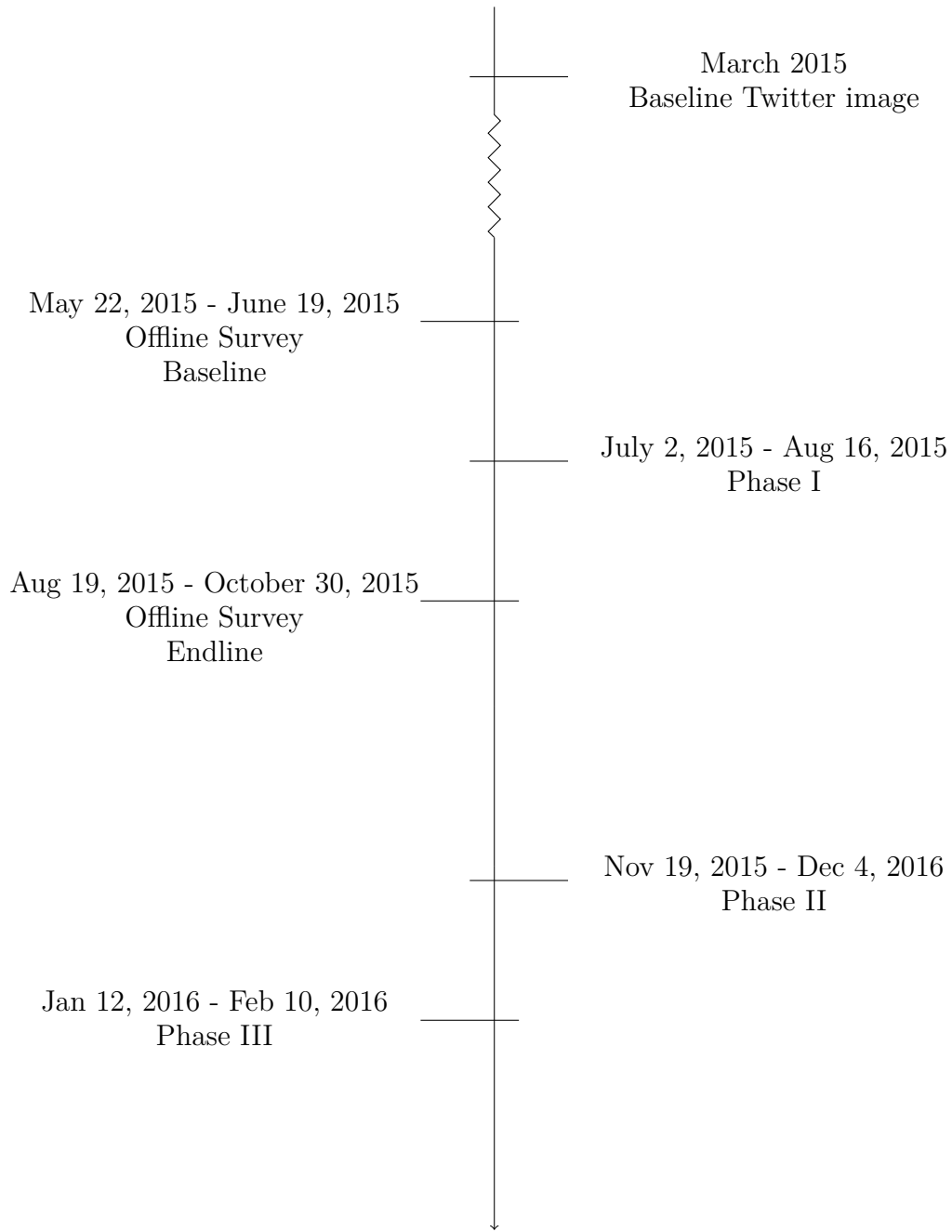
\begin{figure}[!h]
\begin{tikzpicture}[every text node part/.style={align=center}]
\def \n {5}
\def \radius {2cm}
\def \margin {8}

\draw[<-] (0,-2) -- (0,3.2);
\draw (0,3.2) -- (0,5.5);
\draw (0,5.5) -- (0,12.2);
\draw[snake] (0,12.2) -- (0,14);
\draw (0,14) -- (0,15);
\draw(0,15) -- (0,15.5);

\foreach \y in {3, 9, 14.5} 
      \draw (30pt,\y cm) -- (-10pt,\y cm);
      
\foreach \y in {1,7,11}
      \draw (10pt,\y cm) -- (-30pt,\y cm);

    \draw (-4,0.5) node[below=0pt] {$  $} node[above=0pt] {Jan 12, 2016 - Feb 10, 2016\\ Phase III};
    \draw (4,3.5) node[below=0pt] {Nov 19, 2015 - Dec 4, 2016 \\ Phase II} node[above=3pt] {$  $};
    \draw (-4,6) node[below=0pt] {$  $} node[above=0pt] {Aug 19, 2015 - October 30, 2015 \\ Offline Survey \\ Endline};
    \draw (4,9.5) node[below=0pt] { July 2, 2015 - Aug 16, 2015 \\ Phase I} node[above=0pt] { };
    \draw (-4,10) node[below=0pt] {$  $} node[above=0pt] {May 22, 2015 - June 19, 2015 \\ Offline Survey \\ Baseline };
    \draw (4,15) node[below=0pt] {March 2015 
      \\ Baseline Twitter image} node[above=0pt] { };

\end{tikzpicture}
\caption{Timeline\label{fig:timeline}}
\end{figure}

Second, to measure the overall effect of exposure, we randomized our celebrities into two groups, stratified by number of followers (above or below median), style, celebrity or organization, and tweet count. 
Group I celebrities tweeted in the first phase of the experiment (July and August 2015), while Group II celebrities tweeted in the second and third phases of the experiment (November 2015 - February 2016) (see Figure \ref{fig:timeline}). The key point is that our offline endline survey (see Section \ref{subsec:data}) was conducted \emph{in between} these two phases (i.e., late August - October 2015). This means that, conditional on the number of celebrities in our study a given Twitter user follows, at the time of the endline survey it is random how many of these celebrities have actually tweeted about immunization. We use this between-celebrity randomization to estimate the impact of the Twitter campaign on offline beliefs in Section \ref{sec:offline} below.

Summarizing, the experimental design is as follows:
\begin{itemize}
	\item Every celebrity and organization is randomly assigned either to Phase I or Phases II/III.
	\item Conditional on being active in a phase, each celebrity is active on a given day with probability that corresponds to the number of tweets/retweets the celebrity or organization signed up for being randomly distributed over the days in the assigned phase. 
	\item Conditional on being active in a given day, each celebrity tweets one time. 
	\item Each tweet event on a given day is then randomly assigned a time from the empirical distribution of tweets by day-of-week we have from historical Twitter data in Indonesia, constrained between the hours of 7 a.m. to 11 p.m. Jakarta time.
	\item Conditional on a given tweet/retweet event, it is randomly assigned with probability 1/2 to be an celebrity-composed tweet and with probability 1/2 to be a Joe/Jane-composed tweet with the celebrity or organization retweeting it.
	\item Each of these events is then randomly drawn a tweet from the tweet bank, conditional on the organization or celebrity's style.
	\item For a subset of tweets, they were randomized to receive a source.\footnote{Another set of tweets always carried a source because they covered more sensitive material. They were also given priority to be sent out, so they comprise a larger share of low tweet count celebrities' tweets. Consequently in analyzing source citation, we condition appropriately on non-sensitive tweets, to ensure we only compare non-sensitive material tweets which would be randomly assigned to receive a source.}
	
\end{itemize}

\subsection{Data}\label{subsec:data}
We use two types of data in the study, \emph{Online Data}, (i.e., data collected via Twitter about behavior on the Twitter platform with respect to our tweets), and \emph{Offline Data}, (i.e., data collected via a phone survey about respondents' beliefs). We discuss these in turn.

\subsubsection{Online data}\label{subsec:onlinedata}
We collected detailed data on relevant behavior on the Twitter platform via the Twitter Firehose and Twitter API. Before the experiment began, in early 2015, we collected an image of the publicly available Twitter network, including the list of followers of any celebrity participating in our study. 
  This allows us to construct the follower network in Indonesia at baseline. 

There are two main types of behaviors that people who see tweets can do, ``likes'' and ``retweets''. A \emph{like} is a expression of approval of the tweet. A \emph{retweet} is when someone who has seen a tweet rebroadcasts it to their entire follower network; this allows information to propagate over the Twitter network. There are two main differences between likes and retweets. First, retweets do not necessarily imply endorsement of the views of a tweet, whereas likes do. Second, while likes are visible (a user can look up which tweets another user has liked, and can look up who has liked a given tweet), likes are not automatically pushed out as tweets to a user's followers. An example of a campaign tweet, with a source, is depicted in Figure \ref{fig:example}. 

For each the of the 672 total tweets that were originated by our experiment, we tracked each time the tweet was liked or retweeted by any of the over 5.5 million unique users who followed   at least one of the participants in our study.  
  When the tweet was retweeted by a celebrity's follower, we also scraped all of this follower's followers and their liking and retweeting behavior.\footnote{Since a given user can follow multiple celebrities, the 7.8 million total followers of celebrities in our sample 
 	 represents 5.5 million unique users.} For each of these events, we used the complete follower network (and followers' followers when the follower retweeted) from the baseline to construct the shortest path through which the tweet could have reached the user.  
    We denote those retweets / likes coming from a direct follower of a celebrity as $F_1$ events, and those retweets / likes coming from a follower of a follower of a celebrity as $F_2$ events. We use the distinction between $F_1$ and $F_2$ events in more detail in the analysis below.

Table \ref{tab:user_summary}, Panel A reports descriptive statistics using Twitter data on our subjects. Our celebrities on average have 262,648 followers whereas the organizations have 145,300 followers. The followers of our celebrities have on average 1,379 followers. The Joes/Janes that we recruited have on average 508 followers. 

\subsubsection{Offline data}\label{offlinedata}
To measure whether online conversations led to offline behavioral changes, we conducted a phone survey on a sample of 2,441  subjects, all of whom followed at least one of our study participants on Twitter. The phone survey was designed to capture information about immunization (including beliefs in some of the various myths our tweets were intended to counteract), immunization history for children in the family and knowledge of recent immunizations of children of close friends, and questions about immunization and Twitter.

To recruit this sample, we advertised with promoted tweets on Twitter a recruitment to participate in a healthcare survey that particularly targeted women.  We specifically targeted the recruitment ads to the 5.5 million unique users who followed participants in our study, with higher rates of ads for women and for individuals who followed more than one study participant. When users came to the website to sign up to participate, we obtained their Twitter handle 
so that we could link them to the Twitter network.
This process resulted in 2,441 total subjects, all of whom we could link to Twitter. All respondents were surveyed by phone during the endline period; we also contacted a subsample of these respondents (approximately 73 percent) by phone for a baseline survey prior to the beginning of Phase I tweets.

Table \ref{tab:user_summary}, Panel B reports demographics of our offline survey sample. To gauge the sample selection in our sample, we also present comparable data from the 2014 wave of the SUSENAS, the annual representative Indonesian national household survey. Relative to the nationally representative SUSENAS sample, we see that our demographic is more urban, slightly younger, and have a similar gender composition.

Panel C reports baseline statistics for beliefs about vaccinations. We see that there is considerable confusion about the nature and value of vaccines. For instance, only 56 percent  of individuals thought that vaccines are domestically made (they are), and only 38.5 percent  thought that vaccines are free of cost (they are). This suggests substantial room for improvement on immunization knowledge in our study sample.

\

\section{The Value of Endorsement}\label{sec:value_endorsement}

This section explores our main results that characterize the value of endorsement per se. The key idea is that if a celebrity is known to be involved in passing a message, or have authored a message, this provides scope for an endorsement effect. Similarly, a message explicitly listing a source may be more likely to be passed on if viewed as more credible, whereas if it lacks a source, this may impede transmission. On the other hand, source inclusion may impede transmission, say if passing such a benefit accrues the sender less social value.  Further, multiple exposures as compared to a single exposure may lead to more passing of information. 

To untangle these mechanisms, in this section we use the variation we induced at the  tweet level---i.e., the randomly induced variation in who tweeted or retweeted what when. Since the variation we explore here is much finer, our outcome variables must also be at the level of the individual tweet.\footnote{In particular, the randomizations we exploit here are within-celebrity, and hence within-Phase. Therefore we cannot use other tweet-level randomizations in this section to look at  offline beliefs or behavior.} We therefore focus in this section on online responses---``likes" and ``retweets" by ordinary Twitter users of messages composed or shared by our celebrities---that allow us to differentiate responses both tweet by tweet, and also differentially depending on how a particular tweet reached a particular user. 

\subsection{Value of celebrity involvement}\label{subsec:value_endorsement}

We begin by asking whether a celebrity's influence in diffusion on social media is due simply to their reach (the size of their network) or also because when they pass on it is more likely to be subsequently passed on (the endorsement effect due to their involvement).

Our identification strategy exploits a particular feature of retweets in Twitter. A respondent $j$ who sees a retweet observes two names: the name of the original writer of the tweet, and of the person whom $j$ follows who retweeted it. Any names in between---say, a follower of the original writer who retweeted it to the person who retweeted it to $j$---are unobserved.

\begin{figure}[!h]
\centering
\vspace{1cm}
\subfloat[Message $M$ originated by Joe]{
\begin{tikzpicture}
\def \n {5}
\def \radius {2cm}
\def \margin {8}
\node[draw, circle, fill = blue!45, minimum size=19pt] at (-1,0) (v1){$Joe$};
\node[draw, circle, fill = blue!25, minimum size=19pt] at (1,0) (v2){$Celeb$};
\node[draw, circle, fill = blue!25, minimum size=19pt] at (3,0) (v3){$F_1$};
\node[draw, circle, fill = blue!25, minimum size=19pt] at (5,0) (v4){$F_2$};

\draw[line width = 0.3mm,->, >=latex] (v1) to (v2);
\draw[line width = 0.3mm,->, >=latex] (v2) to (v3);
\draw[line width = 0.3mm,->, >=latex] (v3) to (v4);

\end{tikzpicture}
}
\subfloat[$F_2$'s observation]{
\includegraphics[scale = 0.6]{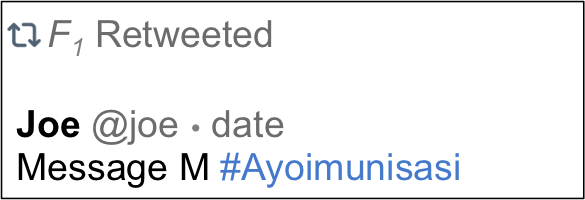}
}

\subfloat[Message $M$ originated by Celeb]{
\begin{tikzpicture}
\def \n {5}
\def \radius {2cm}
\def \margin {8}
\node[draw, circle, , minimum size=19pt] at (-1,0) (v1){$Joe$};
\node[draw, circle, fill = blue!45, minimum size=19pt] at (1,0) (v2){$Celeb$};
\node[draw, circle, fill = blue!25, minimum size=19pt] at (3,0) (v3){$F_1$};
\node[draw, circle, fill = blue!25, minimum size=19pt] at (5,0) (v4){$F_2$};

\draw[line width = 0.3mm,->, >=latex] (v2) to (v3);
\draw[line width = 0.3mm,->, >=latex] (v3) to (v4);

\end{tikzpicture}
}
\subfloat[$F_2$'s observation]{
\includegraphics[scale = 0.6]{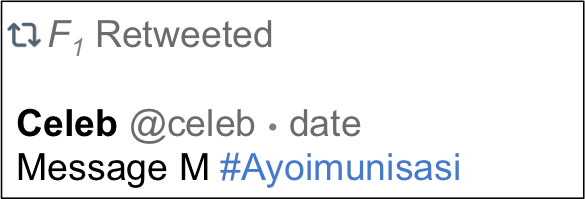}
}
\caption{Identification of the value of endorsement of celebrity involvement.\label{fig:reach-vs-endorsement}}
\end{figure}
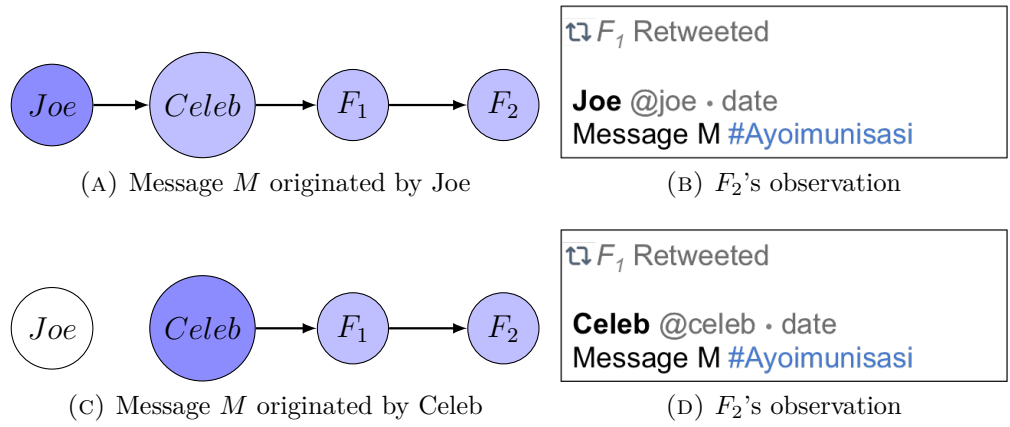

We exploit this feature of Twitter---that intermediaries in a retweet chain are lost---in our experimental design, as summarized in Figure \ref{fig:reach-vs-endorsement}. Consider a chain from a celebrity to some follower $F_1$ and then to some follower of this follower (who does not directly follow the celebrity) $F_2$. If the celebrity retweets the message by a Joe/Jane, and then this is retweeted by $F_1$, observe that $F_2$ sees the message, sees that it is composed by a Joe/Jane, and knows that $F_1$ retweeted it. But crucially $F_2$ does not know that the celebrity had retweeted it: $F_2$ is likely to be blind to the celebrity's involvement. On the other hand, if the celebrity had written this tweet herself rather than retweeted it, this would be visible to the $F_2$. This is depicted in Figure \ref{fig:reach-vs-endorsement}.

By randomizing whether the message is originally tweeted by the celebrity, or instead originally tweeted by a Joe/Jane and then retweeted by the celebrity, we can identify the celebrity endorsement effect by looking at $F_2$'s behavior.

To test this, we estimate, by Poisson regression, the equation 
\begin{equation}
\E[y_{trcmp}\vert {\bf{x}}_{trcmp}] = \exp\left(\alpha\cdot \text{Celeb}_{tcm} + \beta \cdot \text{log(Followers)}_r + \omega_c + \omega_m  + \omega_p\right)
\label{eq:reach_endorse_estimating}
\end{equation}
where $t$ indexes a tweet, $r$ indexes a retweeter (i.e., an $F_1$ who retweeted the tweet $t$), $c$ indexes a celebrity,  $m$ indexes the type of message content, and $p$ indexes phase. The variable $\text{Celeb}_{tcm}$ is a dummy that takes 1 if the celebrity authored the tweet herself (and hence her identity if visible to the $F_2$), and 0 if the celebrity retweeted a Joe/Jane (and hence her identity is not visible to the $F_2$). Each observation is a retweet of one of our original tweets, and the dependent variable $y_{trcmp}$ is a count of how many times this retweet was itself either liked or retweeted again by an $F_2$. Since $y$ is a count, we estimate a Poisson regression, with robust standard errors to allow for arbitrary variance terms clustered at the original tweet ($t$) level. 
We control for the log number of followers of the $F_1$, and for dummies ($\omega_m$) for the different types of messages (e.g., dummies for it being about a fact, importance of immunization, etc). All regressions include celebrity fixed effects ($\omega_c$), which absorb variation   casual/formal style chosen, etc., as well as phase fixed effects ($\omega_p$). Standard errors are clustered at the level of original tweet $t$, which is the level at which $\text{Celeb}_{tcm}$ is randomized. 

The key coefficient of interest is $\alpha$, which measures the differential impact of the tweet having been written by the celebrity (as compared to being written by a Joe/Jane) and this being observable to the $F_2$-level person making the decision to retweet.  

Table \ref{tab:celeb_v_org_v_joe_F2} presents our results. As discussed in more detail in Section \ref{subsec:onlinedata} above, we have three main outcome variables: (1) whether the agent either liked or retweeted the tweet,  (2) whether an agent liked the tweet, and (3) whether an agent retweeted the tweet.  Columns 1, 3, and 5 present the results on the full sample for each of these dependent variables.

We see large endorsement effects. Having a celebrity compose and tweet the message relative to having a Joe/Jane compose the message and the celebrity retweeting it leads to a 1.7-fold increase in the retweet or like rate (column 1, $p = 0.001$; note that since this is a Poisson model, the coefficients are interpretable as the change in log number of retweets) by followers-of-followers ($F_2$s) . The results are similar when we look at likes or retweets alone. 

These results imply that, holding the content of the tweet constant (since it is randomized across tweets) and holding the $F_2$ position in the network constant (since they are all followers-of-followers of the celebrity), having the $F_2$ be aware of the celebrity's involvement in passing the message almost triples the likelihood that the $F_2$ responds online.

We document similar effects of an organization being the originator rather than a Joe/Jane in Table \ref{tab:celeb_v_org_v_joe_F2_appendix} of Appendix \ref{sec:orgs}.\footnote{Recall that we only have 7 organizations, which reduces the overall instances of such cases, so we relegate this to an appendix. Also, we condition on non-sensitive tweets for this sample.} We show that similar to celebrity effect, an organization being randomly assigned to compose a message rather than a Joe/Jane has a substantial endorsement effect of similar magnitude.

There are, however, two main potential threats to our identification strategy here. The first is that when we look at $F_2$ agents, i.e., those who are at distance two from the celebrity of interest, whether a given agent sees a retweet from his or her $F_1$'s may be endogenous and respond to our treatment. Namely, which $F_1$s choose to retweet the message may be directly affected by the fact that the celebrity composed the message, rather than retweeting it from a Joe/Jane. In equation \eqref{eq:reach_endorse_estimating}, we always control for the log number of followers of the $F_1$ who retweeted the message, and hence the number of $F_2$s who could potentially retweet it, so there is no mechanical reason there would be a bias in equation \eqref{eq:reach_endorse_estimating}. But there may nevertheless be a \emph{compositional} difference in which $F_1$'s retweet it, which could potentially lead to selection bias in terms of which $F_2$s are more likely to see the retweet in the first place. 

To address this issue, in Phase III of the experiment, we added an additional randomization. Specifically, in Phase III of the experiment we use the subset of Joe/Jane who are also $F_1$s, and so direct followers of our celebrities. For some of these Joes/Janes, we randomly had their accounts retweet our celebrities' tweets and retweets in the experiment; that is, we created exogenous $F_1s$. For this sample, we can look at how \emph{their} followers---that is, the followers of $F_1$ Joe/Jane's we exogenously forced to retweet a particular tweet---responded as we randomly vary whether the celebrity, an organization, or a Joe/Jane composes the message. We analyze this experiment by estimating equation (\ref{eq:reach_endorse_estimating}) just as we did for the full sample of $F_2$s, but for this sample we have the advantage that whether an $F_2$ sees the tweet is guaranteed to be exogenous by construction.

Columns 2, 4, and 6 present the results. The point estimates are if anything somewhat larger than those in the full sample, and we cannot reject equality.  Statistical significance is reduced somewhat in this restricted sample ($p$-values of 0.119, 0.111, and 0.107 in columns 2, 4, and 6 respectively), but the fact that results are broadly similar to the overall effects in columns 1, 3, and 5 suggests that the possible endogenous selection of $F_1$s in our whole sample is not leading to substantial bias.\footnote{Note that the mean level of retweets in this sample is 0.01, compared to 0.04 in the main sample, which leads to larger standard errors in these columns in a Poisson model. Of course, there is no reason the $F_2$s in this sample would have necessarily have the same retweet rates overall as in the main sample; the point of this exercise is to make sure the $F_2s$ are exogenous to $\text{Celeb}_{tcm}$, but this sample by construction is not meant to be a representative subsample of the entire $F_2$ network.}  

\begin{figure}[!h]

\subfloat[Celebrity Tweet: casual with credibility boost]{
\includegraphics[scale = 0.7]{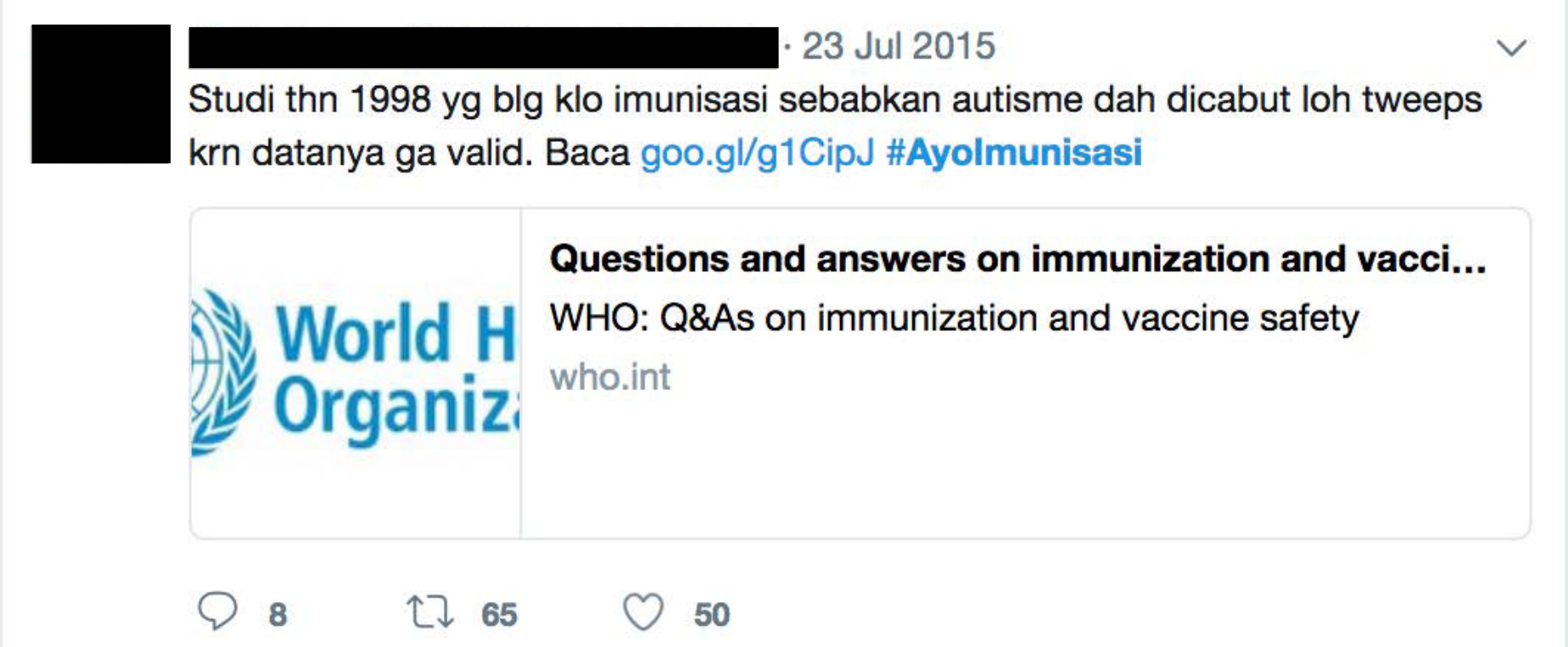}
}
\
\subfloat[Celebrity retweeting an Organization: casual with credibility boost]{
\includegraphics[scale = 0.85]{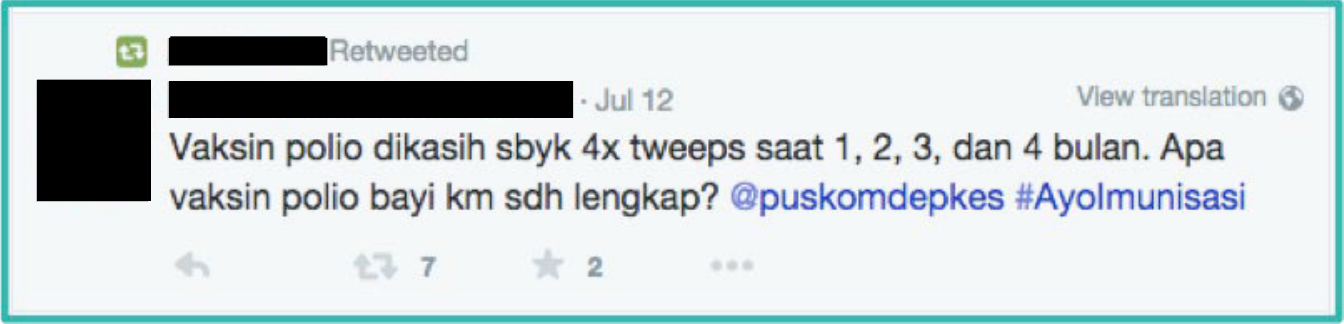}
}
\
\subfloat[Celebrity retweeting a Joe: formal without credibility boost]{
\includegraphics[scale = 0.85]{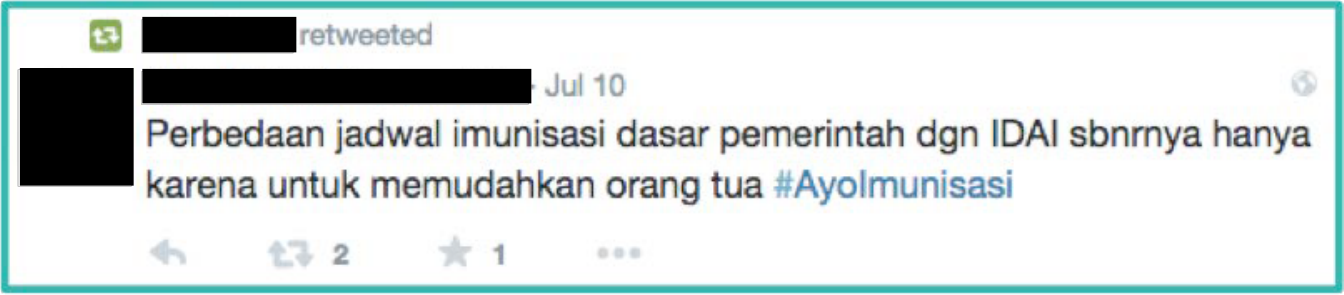}
}

\caption{Sample tweets and retweets from the campaign\label{fig:example}}

\end{figure}

The second potential confound comes from the fact that a retweet carries with it information about how many times the original tweet has been retweeted or liked as of the time the user views it (see Figure \ref{fig:example}, which shows the number of retweets next to the arrow graphic and the number of likes next to the heart graphic). One may worry then that since our treatment assignment affects the retweet count, this itself could spur further changes in the likelihood of retweeting. The same Phase III randomization of forced Joe/Jane retweets also helps us address this issue, because we randomly varied the number of Joes/Janes we forced to retweet a particular tweet.  Appendix \ref{sec:rt_count}, Table \ref{tab:exjoegroup_F1_F2} presents a Poisson regressions of retweets and likes on the number of Joes/Janes that were forced to retweet a given celebrity's tweet or retweet (this is of course net of the forced Joe/Janes' behavior). What we find is that being randomly assigned one, five, ten, or even fifteen extra retweets makes no impact on the number of $F_1$ or $F_2$ retweets that the given tweet faces.

Taken together, our results suggest that an agent who knows that the message was composed by an authority in the sense of a celebrity as compared to thinking it is a random individual is 1.7 times more likely to retweet or like the tweet. This indicates that there is a large endorsement effect.

\subsection{Value of celebrity authorship}

The preceding analysis compared individuals who were effectively randomly blinded to whether a celebrity was or was not involved in the message composition and passing in order to estimate an endorsement effect. We next ask, for a given individual, even if they know that a celebrity is involved and endorsing the message by passing it on, is there an extra endorsement effect from knowing that the celebrity composed it as compared to simply being a conduit?

Table \ref{tab:celeb_v_org_v_joe_F1} presents the results of Poisson regressions at the tweet level. That is, we now restrict to direct followers of the celebrity ($F_1$ individuals), and estimate
\begin{equation}
\E[y_{tcmp}\vert {\bf x}_{tcmp}] = \exp\left(\alpha \cdot \text{Celeb}_{tcm} + \omega_c + \omega_m  + \omega_p\right).
\label{eq:reach_endorse_estimating_f1}
\end{equation}

We now have one observation per tweet, and look at the number of retweets/likes, retweets, or likes by $F_1$ agents who are distance 1 from the celebrity passing along the tweet. We continue to include celebrity ($\omega_c$), phase ($\omega_p$),  and message-type ($\omega_m$) fixed effects. We are interested in whether the tweet being randomly assigned to be composed by the Joes/Janes versus the celebrity leads to a higher retweet rate. 

We find evidence of a large endorsement through authorship effect for celebrities with the retweet/like rate increasing 3-fold (column 1, $p<0.001$). In fact, an agent who observes a tweet composed from the celebrity rather than a retweet of a Joe/Jane is 2.2 times more likely to like the tweet (column 2, $p<0.001$) and 3.8 time more likely to retweet the tweet (column 3, $p < 0.001$).

Again, keeping in mind that we have only 7 organizations, we can look at the same effect of a celebrity retweeting an organization rather than a Joe/Jane in Table \ref{tab:celeb_v_org_v_joe_F1_appendix} of Appendix \ref{sec:orgs}.\footnote{We must condition on non-sensitive topic tweets for this analysis.} We find no added value here: that is, a celebrity retweeting a Joe/Jane and a celebrity retweeting an organization is just as valuable, whereas if the celebrity is known to have composed the message there is a large premium.

On net, the results in this section complement the results in Section \ref{subsec:value_endorsement}, which together paint a consistent picture: celebrity endorsement seems to substantially increase the probability messages are passed on or liked, both overall (as above) and even within the distinction of a celebrity writing vs. retweeting the same message.

\subsection{Value of source citation}

The next question we ask is whether including source citation increases diffusive behavior.  Source citations in our context come in several forms. First, sources are simply provided though there is no verifiable references. This is of the form ``Polio vaccine should be given 4 times at months 1, 2, 3, 4. Are your baby's polio vaccines complete? @puskomdepkes' where ``@puskomdepkes'' is a link to the Twitter handle of the Ministry of Health (known as \textit{DepKes} in Indonesian). Second, explicit sources are cited, and there is a Google shortened link provided.\footnote{Note that Twitter automatically produces a short preview of the content if the site linked to has Twitter cards set up. There is one non-Google shortened link used when citing IDAI (Ikatan Dokter Anak Indonesia, the Indonesian Pediatric Society).}

To examine this question, we re-estimate equation \eqref{eq:reach_endorse_estimating_f1} at the $F_1$ level, but also add a variable that captures whether the tweet was randomly selected to include a source.\footnote{Note that the number of observations is smaller here, because some tweets on topics deemed `sensitive' by the Government always included a source, as we noted above. We restrict the analysis here to those tweets for which we randomized whether the source was included or not.} Table \ref{tab:credboost_F1} presents the results. Columns 1-4 look at pooled likes and retweets, 5-8 look at retweets, and 9-12 look at likes. We also condition the regressions to the sample where the celebrity retweets the Joes/Janes (columns 2, 6, 10), the celebrity directly tweets (columns 3, 7, 11), and the celebrity retweets an organization (columns 4, 8, 12).

On average, pooling across all messaging configurations we find that source citation reduces the retweet and liking rate by 26.3 percent  ($p = 0.051$). This is particularly driven by reduced retweeting behavior (a decline of 27.2 percent, $p = 0.048$). Disaggregating across whether the celebrity composed and tweeted the message or retweeted a Joe/Jane or an organization, we find that the large reductions in retweet rates persist when a Joe/Jane composed the message (a 50 percent decline, $p = 0.02$) or when the celebrity directly tweets the message (a 29.3 percent  decline, $p = 0.002$).
Sources only cease to have an adverse effect and wind up having no effect when an organization itself (here largely health based organizations) originally tweets the message, which is consistent with the idea that the organization itself is essentially interpreted as a source. 

In sum, for both Joes/Janes and celebrities, including a source in tweets ultimately depresses retweet rates and the extent of diffusion. This result---that sources depress retweet rates---may seem surprising, since one might expect that a sourced message may be more reliable. But recall the discussion in Section \ref{sec:design} suggesting that each feature of the message (e.g., originator identity, sourcing) could have ex ante ambiguous effects on retweet rates.  
 There are 
  a number of possible explanations for this finding. One idea is that for an $F_1$ passing on a message has both instrumental value (delivering a good message), as well as a signaling value (conveying to followers that the $F_1$ is able to discern which information is good). We discuss in Appendix \ref{sec:Model} how adding a source could potentially reduce the signaling value and possible lead to lower retweet rates (alongside a more general discussion of how celebrity origination, source citation, exposure, and specific content could all either increase or decrease retweet rates ex ante). 
 Various other stories are possible as well. For instance, it is also possible that $F_1$s interpreted a sourced message as less authentic-sounding than an unsourced message, or perhaps the information sounds less novel when sourced. Regardless, the result suggests that adding explicit credibility-boosts to celebrity messages does not necessarily increase diffusion. 

\subsection{Value of multiple exposures}

Lastly we look at the role of multiple exposures as compared to a single exposure. In models of diffusion there is a contrast between ``simple'' and ``complex'' contagion models, both of which are extensively studied. The distinction matters for the design of campaigns aiming to maximize diffusion.  In simple contagion, adoption happens at the point of a single exposure to the information.  In contrast, there are ``complex contagion'' models, wherein an individual must have multiple exposures before adoption \citep*{centola2007complex,romero2011differences,ghasemiesfeh2013complex}.\footnote{A micro-foundation for complex contagion-like behavior is as follows:  consider a Bayesian individual who is unsure about whether it is worthwhile to tweet or retweet about immunization. Her default is to not tweet: she only tweets or retweets if she is sure enough that it is worthwhile, because say there is a small cost to passing on a message that is not worth it. In this case, she only acts when she is sure enough; with enough exposures she may clear the threshold and then act, and at that point a marginal exposure should not matter.}  \cite*{beaman2016can} provide evidence in favor of complex contagion in a different setting---that of technological adoption in rural networks. 
While the networks literature across myriad disciplines (e.g., computer science, applied math, sociology, economics) have different perspectives on why this may be the case, roughly speaking complex contagion processes differ from simple contagion processes in that there is a premium for identifying central individuals. This makes it considerably more likely that a given individual in the network receives multiple exposures and thereby affects the odds of something going viral.

To get at these issues, we look at whether there are linear or non-linear effects to exposure to a message.   
Table \ref{tab:panel_daylevel_exposure} presents a Poisson regression to examine the shape of the response function. We look at all agents at the $F_1$ level, who are followers of our celebrity in question, and the observation is at the user-date level. We ask how the number of retweets of our campaign's tweets that they see from celebrities they follow on that day by the agent depends on the (randomly assigned) exposure to tweets on the given day. Since celebrities are only assigned to tweet once per day, multiple signals will come from users who follow multiple celebrities. Further, because we are interested in non-linear effects, we condition on a sample who follow at least 3 celebrities. We estimate a regression of the form

\begin{equation}
\E [y_{id} \vert {\bf x}_{id}] = \exp \left( \alpha + \sum_{k=2}^K \beta_k {\bf 1}\{\text{Exposure to }k\text{ tweets today}_{id}\} + \mu_i + \mu_d \right),
\label{eq:exposure}
\end{equation}
where $y_{id}$ is the number of retweets that an $F_1$ individual $i$ makes on date $d$ out of the set of campaign tweets that he is directly exposed to from celebrities he follows and which includes user and date fixed effects.

Going from 1 exposure to 2 exposures leads to just over a two-fold increase in the retweet rate or a 110 percent  
increase in the retweet rate ($p = 0.01$). That there is an effect suggests that we are not in a world of simple contagion. Then going from 1 to 3 corresponds to a 2.6-fold or 160 percent increase in the retweet rate ($p = 0.08$).  Finally going to 4-7 exposures leads to no differential retweet rate relative to 1, though the point estimate is large but noisily estimated. This clearly rejects a linear effect model: the first additional exposure contributes to the largest increase in the retweet rate, the second additional tweet has additional value,  and there is no effect for marginal exposures, which indicates concavity.

On net, the evidence suggests the retweet shape function begins linearly and then flattens out. This is inconsistent with simple contagion. But moreover, notice this is also inconsistent with a model of attention where an agent randomly checks her phone and then decides to retweet, since in this case the retweet count should scale linearly, but in fact it is concave.

Ultimately, the evidence is consistent with a complex contagion-like model that having multiple tacit endorsements is enough to make it worthwhile for the user to behave. This is particularly important because the theory demonstrates that there is a striking difference in whether a policymaker must strive to contact the ``right'' individuals (central individuals) or not---complex contagion models have a premium in seeding information with influential. The data then shows that even without the endorsement effect itself, seeding influential individuals with information would have more value simply through the fact that it would generate more multiple exposures.

\subsection{Discussion}

In this section we have studied how endorsement effects of various kinds may influence consumption of and passing on information. The results are nuanced. Our design allows us to parse the endorsement effect of a celebrity from the reach effect, and using this approach, we find the endorsement effect per se to be large: both likes and retweets increase considerably when a celebrity is randomly assigned to write the same message as a Joe/Jane. At the same time, inclusion of credible sourcing in the message does not help: on average it reduces retweeting. Further exposures themselves do matter, contrary to common views on diffusion, but in a concave way. Multiple exposures help at the outset but subsequently decline in marginal value.

This suggests certain important themes to focus on in future studies of diffusion. First, more attention ought to be paid to  ``node identity''---that is, who is involved in writing or passing the information and at what stage. This plays an enormous role in our setting, as having the celebrity be the source plays an outsized role per se. This type of concern is often swept aside, due to the obvious complexities, in typical studies of diffusion. A notable exception is \cite{beaman2018diffusion}, who focus on how the gender of the individual in the network strongly affects diffusion of information. Second, the results suggest certain types of diffusion are more likely: in this case multiple exposures (closer in spirit to complex contagion) plays a crucial role, in keeping with recent work in other contexts (e.g., \cite{beaman2016can} for technology adoption). Third, the extent of diffusion itself, typically parametrized either by time or the intrinsic virality of the information, may greatly depend on the topic. While taken for granted, there is limited research in terms of how we expect diffusion to differ by content-type and how this interacts with both the \emph{who} and the \emph{how much}.

\

\section{Does Online Discussion Have Offline Effects?}\label{sec:offline}

Next we ask whether an online celebrity endorsement campaign can begin to have measurable offline effects. To investigate this, we use the fact that between Phases I and II, we conducted an offline phone survey, as described in Section \ref{sec:experiment}, so that conditional on the number of our celebrities a user followed, exposure to our campaign as of the time of the phone survey was randomly assigned.

\begin{figure}[!h]
\subfloat[Twitter network with offline survey nodes denoted by diamond]{
\scalebox{0.8}{
\begin{tikzpicture}
\def \n {5}
\def \radius {2cm}
\def \margin {8}

\node[draw, circle,  minimum size=12pt] at (-2,0) (v1){$Celeb$};
\node[draw, circle,  minimum size=12pt] at (-2.1,-1.9) (v2){$$};
\node[draw, diamond,  minimum size=12pt] at (-4,-1) (v3){$$};
\node[draw, circle,  minimum size=12pt] at (-3.2,1.6) (v4){$$};
\node[draw, circle,  minimum size=12pt] at (-2,2) (v5){$$};
\node[draw, diamond,  minimum size=12pt] at (0,1.5) (v6){$$};
\node[draw, diamond,  minimum size=12pt] at (1,0.2) (v7){$$};
\node[draw, diamond,  minimum size=12pt] at (0.5,-1) (v8){$$};

\draw[line width = 0.3mm, >=latex] (v1) to (v2);
\draw[line width = 0.3mm, >=latex] (v1) to (v3);
\draw[line width = 0.3mm, >=latex] (v1) to (v4);
\draw[line width = 0.3mm, >=latex] (v1) to (v5);
\draw[line width = 0.3mm, >=latex] (v1) to (v6);
\draw[line width = 0.3mm, >=latex] (v1) to (v7);
\draw[line width = 0.3mm, >=latex] (v1) to (v8);

\node[draw, circle,  minimum size=12pt] at (-1.5,-4) (v9){$Celeb$};
\node[draw, circle,  minimum size=12pt] at (-2.3,-5.9) (v10){$$};
\node[draw, diamond,  minimum size=12pt] at (-4.2,-5) (v11){$$};
\node[draw, circle,  minimum size=12pt] at (-3,-3.9) (v12){$$};
\node[draw, diamond,  minimum size=12pt] at (-1.8,-2.7) (v13){$$};
\node[draw, circle,  minimum size=12pt] at (0,-2.5) (v14){$$};
\node[draw, diamond,  minimum size=12pt] at (1.2,-3.8) (v15){$$};
\node[draw, circle,  minimum size=12pt] at (0.5,-5) (v16){$$};

\draw[line width = 0.3mm, >=latex] (v9) to (v10);
\draw[line width = 0.3mm, >=latex] (v9) to (v11);
\draw[line width = 0.3mm, >=latex] (v9) to (v12);
\draw[line width = 0.3mm, >=latex] (v9) to (v13);
\draw[line width = 0.3mm, >=latex] (v9) to (v14);
\draw[line width = 0.3mm, >=latex] (v9) to (v15);
\draw[line width = 0.3mm, >=latex] (v9) to (v16);

\draw[line width = 0.3mm, >=latex] (v2) to (v13);
\draw[line width = 0.3mm, >=latex] (v3) to (v13);
\draw[line width = 0.3mm, >=latex] (v6) to (v14);
\draw[line width = 0.3mm, >=latex] (v7) to (v15);

\node[draw, circle,  minimum size=12pt] at (6,0) (v17){$Celeb$};
\node[draw, circle,  minimum size=12pt] at (5.9,-1.9) (v18){$$};
\node[draw, diamond,  minimum size=12pt] at (4,-1) (v19){$$};
\node[draw, circle,  minimum size=12pt] at (4.8,1.6) (v20){$$};
\node[draw, diamond,  minimum size=12pt] at (6,2) (v21){$$};
\node[draw, diamond,  minimum size=12pt] at (8,1.5) (v22){$$};
\node[draw, circle,  minimum size=12pt] at (9,0.2) (v23){$$};
\node[draw, circle,  minimum size=12pt] at (8.5,-1) (v24){$$};

\draw[line width = 0.3mm, >=latex] (v17) to (v18);
\draw[line width = 0.3mm, >=latex] (v17) to (v19);
\draw[line width = 0.3mm, >=latex] (v17) to (v20);
\draw[line width = 0.3mm, >=latex] (v17) to (v21);
\draw[line width = 0.3mm, >=latex] (v17) to (v22);
\draw[line width = 0.3mm, >=latex] (v17) to (v23);
\draw[line width = 0.3mm, >=latex] (v17) to (v24);

\node[draw, circle,  minimum size=12pt] at (6.2,-4) (v25){$Celeb$};
\node[draw, circle,  minimum size=12pt] at (6,-5.9) (v26){$$};
\node[draw, diamond,  minimum size=12pt] at (4,-5) (v27){$$};
\node[draw, circle,  minimum size=12pt] at (5.1,-3.9) (v28){$$};
\node[draw, diamond,  minimum size=12pt] at (6.4,-2.7) (v29){$$};
\node[draw, circle,  minimum size=12pt] at (7.8,-2.5) (v30){$$};
\node[draw, diamond,  minimum size=12pt] at (9,-3.8) (v31){$$};
\node[draw, circle,  minimum size=12pt] at (8.2,-5) (v32){$$};

\draw[line width = 0.3mm, >=latex] (v25) to (v26);
\draw[line width = 0.3mm, >=latex] (v25) to (v27);
\draw[line width = 0.3mm, >=latex] (v25) to (v28);
\draw[line width = 0.3mm, >=latex] (v25) to (v29);
\draw[line width = 0.3mm, >=latex] (v25) to (v30);
\draw[line width = 0.3mm, >=latex] (v25) to (v31);
\draw[line width = 0.3mm, >=latex] (v25) to (v32);

\draw[line width = 0.3mm, >=latex] (v19) to (v28);
\draw[line width = 0.3mm, >=latex] (v23) to (v29);
\draw[line width = 0.3mm, >=latex] (v23) to (v31);
\draw[line width = 0.3mm, >=latex] (v24) to (v32);

\draw[line width = 0.3mm, >=latex] (v6) to (v20);
\draw[line width = 0.3mm, >=latex] (v14) to (v19);

\draw[line width = 0.3mm, >=latex] (v3) to (v9);
\draw[line width = 0.3mm, >=latex] (v6) to (v9);
\draw[line width = 0.3mm, >=latex] (v8) to (v17);
\draw[line width = 0.3mm, >=latex] (v7) to (v25);
\draw[line width = 0.3mm, >=latex] (v19) to (v25);

%
%
%
%

\end{tikzpicture}
}
}

\subfloat[Phase I treatment celebrities and treated nodes and offline survey respondents colored in blue. Offline survey respondents beliefs affected by treatment measured by comparing blue and white diamonds.]{
\scalebox{0.8}{
\begin{tikzpicture}
\def \n {5}
\def \radius {2cm}
\def \margin {8}

\node[draw, circle, fill=blue!60,  minimum size=12pt] at (-2,0) (v1){$Celeb$};
\node[draw, circle, fill=blue!35,   minimum size=12pt] at (-2.1,-1.9) (v2){$$};
\node[draw, diamond, fill=blue!35,   minimum size=12pt] at (-4,-1) (v3){$$};
\node[draw, circle, fill=blue!35,   minimum size=12pt] at (-3.2,1.6) (v4){$$};
\node[draw, circle, fill=blue!35,   minimum size=12pt] at (-2,2) (v5){$$};
\node[draw, diamond, fill=blue!35,   minimum size=12pt] at (0,1.5) (v6){$$};
\node[draw, diamond, fill=blue!35,   minimum size=12pt] at (1,0.2) (v7){$$};
\node[draw, diamond, fill=blue!35,   minimum size=12pt] at (0.5,-1) (v8){$$};

\draw[line width = 0.3mm, >=latex] (v1) to (v2);
\draw[line width = 0.3mm, >=latex] (v1) to (v3);
\draw[line width = 0.3mm, >=latex] (v1) to (v4);
\draw[line width = 0.3mm, >=latex] (v1) to (v5);
\draw[line width = 0.3mm, >=latex] (v1) to (v6);
\draw[line width = 0.3mm, >=latex] (v1) to (v7);
\draw[line width = 0.3mm, >=latex] (v1) to (v8);

\node[draw, circle,  fill=blue!60,  minimum size=12pt] at (-1.5,-4) (v9){$Celeb$};
\node[draw, circle,  fill=blue!35,  minimum size=12pt] at (-2.3,-5.9) (v10){$$};
\node[draw, diamond,  fill=blue!35,  minimum size=12pt] at (-4.2,-5) (v11){$$};
\node[draw, circle,  fill=blue!35,  minimum size=12pt] at (-3,-3.9) (v12){$$};
\node[draw, diamond,  fill=blue!35,  minimum size=12pt] at (-1.8,-2.7) (v13){$$};
\node[draw, circle,  fill=blue!35,  minimum size=12pt] at (0,-2.5) (v14){$$};
\node[draw, diamond,  fill=blue!35,  minimum size=12pt] at (1.2,-3.8) (v15){$$};
\node[draw, circle,  fill=blue!35,   minimum size=12pt] at (0.5,-5) (v16){$$};

\draw[line width = 0.3mm, >=latex] (v9) to (v10);
\draw[line width = 0.3mm, >=latex] (v9) to (v11);
\draw[line width = 0.3mm, >=latex] (v9) to (v12);
\draw[line width = 0.3mm, >=latex] (v9) to (v13);
\draw[line width = 0.3mm, >=latex] (v9) to (v14);
\draw[line width = 0.3mm, >=latex] (v9) to (v15);
\draw[line width = 0.3mm, >=latex] (v9) to (v16);

\draw[line width = 0.3mm, >=latex] (v2) to (v13);
\draw[line width = 0.3mm, >=latex] (v3) to (v13);
\draw[line width = 0.3mm, >=latex] (v6) to (v14);
\draw[line width = 0.3mm, >=latex] (v7) to (v15);

\node[draw, circle,  minimum size=12pt] at (6,0) (v17){$Celeb$};
\node[draw, circle,  minimum size=12pt] at (5.9,-1.9) (v18){$$};
\node[draw, diamond,  minimum size=12pt] at (4,-1) (v19){$$};
\node[draw, circle,  minimum size=12pt] at (4.8,1.6) (v20){$$};
\node[draw, diamond,  minimum size=12pt] at (6,2) (v21){$$};
\node[draw, diamond,  minimum size=12pt] at (8,1.5) (v22){$$};
\node[draw, circle,  minimum size=12pt] at (9,0.2) (v23){$$};
\node[draw, circle,  minimum size=12pt] at (8.5,-1) (v24){$$};

\draw[line width = 0.3mm, >=latex] (v17) to (v18);
\draw[line width = 0.3mm, >=latex] (v17) to (v19);
\draw[line width = 0.3mm, >=latex] (v17) to (v20);
\draw[line width = 0.3mm, >=latex] (v17) to (v21);
\draw[line width = 0.3mm, >=latex] (v17) to (v22);
\draw[line width = 0.3mm, >=latex] (v17) to (v23);
\draw[line width = 0.3mm, >=latex] (v17) to (v24);

\node[draw, circle,  minimum size=12pt] at (6.2,-4) (v25){$Celeb$};
\node[draw, circle,  minimum size=12pt] at (6,-5.9) (v26){$$};
\node[draw, diamond,  minimum size=12pt] at (4,-5) (v27){$$};
\node[draw, circle,  minimum size=12pt] at (5.1,-3.9) (v28){$$};
\node[draw, diamond,  minimum size=12pt] at (6.4,-2.7) (v29){$$};
\node[draw, circle,  minimum size=12pt] at (7.8,-2.5) (v30){$$};
\node[draw, diamond,  minimum size=12pt] at (9,-3.8) (v31){$$};
\node[draw, circle,  minimum size=12pt] at (8.2,-5) (v32){$$};

\draw[line width = 0.3mm, >=latex] (v25) to (v26);
\draw[line width = 0.3mm, >=latex] (v25) to (v27);
\draw[line width = 0.3mm, >=latex] (v25) to (v28);
\draw[line width = 0.3mm, >=latex] (v25) to (v29);
\draw[line width = 0.3mm, >=latex] (v25) to (v30);
\draw[line width = 0.3mm, >=latex] (v25) to (v31);
\draw[line width = 0.3mm, >=latex] (v25) to (v32);

\draw[line width = 0.3mm, >=latex] (v19) to (v28);
\draw[line width = 0.3mm, >=latex] (v23) to (v29);
\draw[line width = 0.3mm, >=latex] (v23) to (v31);
\draw[line width = 0.3mm, >=latex] (v24) to (v32);

\draw[line width = 0.3mm, >=latex] (v6) to (v20);
\draw[line width = 0.3mm, >=latex] (v14) to (v19);

\draw[line width = 0.3mm, >=latex] (v3) to (v9);
\draw[line width = 0.3mm, >=latex] (v6) to (v9);
\draw[line width = 0.3mm, >=latex] (v8) to (v17);
\draw[line width = 0.3mm, >=latex] (v7) to (v25);
\draw[line width = 0.3mm, >=latex] (v19) to (v25);

%
%
%
%
%
%
%

\end{tikzpicture}
}
}
\caption{Offline impact randomization schematic\label{fig:slacktivism}}
\end{figure}
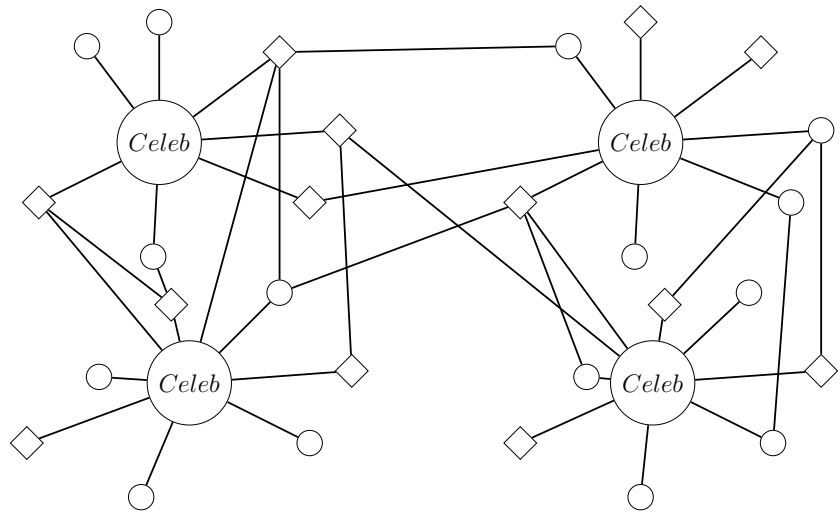
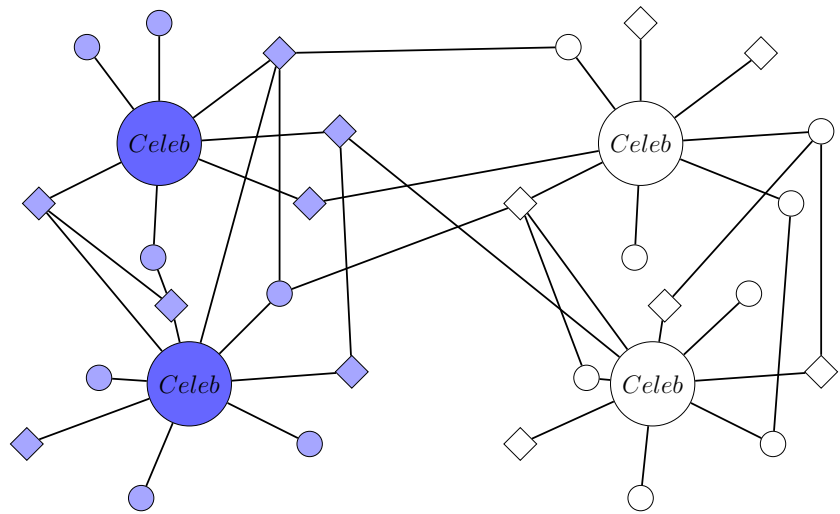

Figure \ref{fig:slacktivism} presents a schematic of the identification. Because those who followed Phase I celebrities faced greater campaign exposure, and celebrities and organizations were randomly binned into Phase I versus II/III, we can look at how exposure to the campaign in terms of tweets at the end of Phase I affects knowledge and behavior. Note that because users who follow more of our celebrities are mechanically more likely to follow people who were in Phase I, we always condition on the potential exposure, i.e., the number of celebrities in our study they follow and how many tweets those celebrities agreed to send as part of the campaign. 
 We thus exploit the random variation in actual Phase I exposure conditional on potential Phase I exposure.  

\subsection{Did people hear about the campaign?}

We begin with what can be thought of as akin to a first-stage in Table \ref{tab:heard_campaign}. We ask whether respondents were more likely to have heard of our hashtag (\#AyoImunisasi) or heard about immunization discussions from Twitter if they were randomly more exposed to campaign tweets, conditional on their potential exposure. In particular, given a location in the network, each respondent had a \emph{potential exposure} under our campaign, which was a function of how many tweets/retweets were assigned to each celebrity they followed. Relative to this, the \emph{actual exposure} uses the fact that because we randomized which celebrities went in Phase I and which in Phases II-III, there was random variation in actual exposure relative to potential exposure when we survey between Phases I and II-III.

We run logistic and Poisson regressions are of the form
\begin{equation}
	f(y_{i}) = \alpha + \beta \cdot \text{Exposure to Tweets}_{i} + \gamma \cdot \text{Potential Exposure}_i + \delta' X_{i},
\label{eq:exposure}
\end{equation}
where $y_i$ is either whether the agent has heard of our hashtag \#AyoImunisasi, whether the agent has heard of immunization in general from Twitter, or the number of times they have heard about immunization from Twitter, and $f(\cdot)$ is the appropriate function for logit (log-odds, i.e., $\log \left( \frac{\Pr(y_i = 1 \vert {\bf x}_i)}{1-\Pr(y_i = 1 \vert {\bf x}_i)}\right)$) or Poisson regression (log of expected count, $\log (\E[y_i  \vert {\bf x}_i])$).  $\text{Exposure to Tweets}_i$ is the number of campaign tweets that $i$ is randomized to see through Phase I (normalized to have standard deviation 1).  $\text{Potential Exposure}_i$ is the total number of campaign tweets that $i$ could potentially seeing through the campaign given the celebrities he follows. 
 $X$ are controls, such as the number of celebrities followed by $i$, the log of the number of followers of celebrities by $i$, survey dates, and (in some specifications) demographics and baseline beliefs, selected here and in subsequent regressions 
by double post-LASSO \citep*{belloni2014high,belloni2014inference}.\footnote{Simply put, as we have many potential covariates for which we could control, by employing machine learning (in this case LASSO), we can in a disciplined manner select which controls ought to be used. The double post-LASSO algorithm applies the LASSO to all covariates to select all variables that are sufficiently predictive of the outcome directly (as in a reduced form) and all variables that are sufficiently predictive of the treatment status (as in a first stage), and then the union of these LASSO-selected variables are included as the right controls in our main regression \eqref{eq:exposure}.  Results are qualitatively similar dropping controls and including all controls, though results using LASSO-selected controls have slightly higher power.}  We report standard errors and $p$-values clustered at the level of the combination of celebrities followed; further, because of the complex nature of the potential correlation in $\text{Exposure to Tweets}_i$ across individuals $i$ induced by partial overlap in which celebrities our survey respondents follow, we present randomization-inference $p$-values as well.\footnote{Specifically, we re-run our randomization programs 2,000 times to generate alternative possible configurations of celebrities randomized into Phase I and II/III, fully respecting the stratification and other randomization parameters. We use these alternate randomization results to generate randomization-inference p-values.} The experimental design of randomizing celebrities into phases means that, while individuals $i$ may differ in the number of our celebrities they follow, $\text{Exposure to Tweets}_i$ is random conditional on $\text{Potential Exposure}_i$. 

Columns 1 examines whether the respondent heard of \#AyoImunisasi (and therefore remembers the exact hashtag), column 2 examines whether the respondent heard of a discussion of immunization on Twitter, and column 3 examines the number of times the respondent heard of such discussions. 

We find that a one-standard deviation increase in exposure to the campaign (15 tweets) corresponds to a 16.75 percent 
 increase in the probability that the respondent had heard of our hashtag relative to a mean of 7.7 percent (clustered $p=0.044$, RI $p=0.107$).\footnote{Note that the table reports impacts on log-odds; we report marginal effects in the text, which are interpretable as percent increases. For example, the 16.75 percent increase in column 1 corresponds to an increase in 0.197 in log-odds.} Further, a one-standard deviation increase in exposure corresponds to a  8.3 percent 
  increase in the probability they heard about immunization in general from Twitter relative to a mean of 18.1 percent (clustered $p=0.106$, RI $p=0.046$). Finally looking at the number of times they heard about immunization from Twitter, there is a 11.2 percent increase 
 relative to a mean of 0.322 times (clustered $p=0.044$, RI $p=0.127$).  

Overall, we do see people seem to have noticed the campaign: the randomized increased exposure to our campaign tweets corresponds to more knowledge of our hashtags and being more likely to have heard of immunization on Twitter in the general population.

\subsection{Did people then increase their knowledge about immunization facts?}

Next we ask whether exposure to the campaign led to increased knowledge about immunization. Our survey asks questions about several categories of knowledge. First, we check for knowledge of several common ``myths'' about vaccination that our campaign tried to cover. In particular, we ask whether people know that vaccines are domestically produced, to combat the common rumor in Indonesia that they contain pig products in production (which would make them unacceptable for Muslims, who represent the vast majority of Indonesia's population; domestic products are known to be halal). Second, we ask whether they believe that natural alternatives (breastfeeding, herbal supplements, alternative supplements) replace the need for immunization.  
  Third, we ask whether they are aware that typical symptoms (mild fevers or swelling) are to be expected and not a cause for alarm. The second category we ask about is ``access'' information; in particular, we ask whether they know that it is free to get one's child vaccinated at government health centers. All of these issues were covered in the campaign.   

The fact that we emphasize the role of messages that dispel myths, such as the fact that vaccines are not domestically produced (and therefore may not be halal), is echoed in online behavior. In particular, Table \ref{tab:content_F1} in Appendix \ref{sec:rt_content} shows that that tweets concerning myths diffused more widely than facts that were non-myths, meaning that the exposure would have been more about myth-busting facts. Moreover, myth-dispelling facts comprised 36.7 percent  of all tweets and 82.4 percent of all fact-related tweets sent out (i.e., myths compared to other facts).

  Table \ref{tab:knowledge_immun} presents the results for each of these four categories of information. Column 1 presents whether they understood vaccines were domestically produced.  Columns 2 and 3 check whether respondents understood that the substitutes are invalid and the side-effects are negligible. As described above, these three are the predominant myth-busting facts provided that we asked about in our survey. Column 4 checks whether they understand that immunization is freely provided, which concerns access. We find knowledge effects for our most prominent tweet---domestic production---though not on rumors about substitutability,  side-effects, nor free access.  Seeing 15 campaign tweets in general corresponded to an increase of 5 percent 
 in the probability of correctly answering the domestic question on a base of 57.6 percent (clustered $p=0.042$, RI $p = 0.028$; if we adjust for the fact that 4 questions are asked using a Bonferroni-style adjustment; these p-values would be 0.168 and 0.112, respectively).

\subsection{Knowledge and behavior of network members}

We then look to measurements of knowledge and reported behavior by respondents and those in their neighbor, friend, and relative networks. First we ask whether the respondents' knowledge of immunization status in each type of network increases. Then, given this, we examine the respondent's reports of recent immunization behavior by those in each type of network, as well as own immunization behavior. Looking at networks allows us to paint a picture of a wider set of individuals---after all, having a young child exactly in this age range is a sparse event---but it is important to note that the caveat that this is as reported by the respondent, not observed by us directly. 
   
 \subsubsection{Knowledge about immunization practices of others} 

We begin by asking whether individuals were more likely to know about what their neighbors, friends, and relatives' immunization behavior was, which would be a byproduct of offline conversations since June 2015 to capture the campaign's effect. Panel A of Table \ref{tab:networks} presents the results. Unsurprisingly, the relative network has the highest mean knowledge to begin with (after all, health conversations presumably are very frequent within family), whereas geographic neighbors have the lowest mean. A similar thing is true among friends. In Indonesia, immunizations take place at monthly \textit{posyandu} meetings, which occur each month in each neighborhood (usually hamlets, or \textit{dusun}, in rural areas, and neighborhoods known as \textit{rukun warga}, or \textit{RW}, in urban areas; see \citet*{olken2014should}), so if knowledge would increase, one might expect it to be the knowledge about immunization practices of ones' neighbors. Note that here the sample is restricted to respondents who know friends, relatives, or neighbors with at least one child (ages 0-5) respectively as this is the relevant set, which reduces the survey sample.

In column 1 we see that being exposed to 15 more tweets corresponds to a 5.2 percent  
increase in the probability of knowing the neighbor's status (clustered $p=0.004$, RI $p = 0.088$).   We do not consistently see significant effects on non-neighbor friends. With relatives the point estimates are comparably large, though the estimates are noisier. In any case, this is also consistent with the idea that for the most part, individuals have a higher rate of knowledge of health status of those they have closer relationships with (e.g., relatives) and that this is unlikely to change.\footnote{We also asked the respondents to rate on a scale of 1-4 whether they felt immunization was beneficial,  safe, and important. The median response is a 4 (the maximal score) and respondents overwhelmingly reported this (means are 3.7, 3.4, and 3.8 on the three questions, respectively). The data is essentially top-coded but we present the result in Table \ref{tab:opinion_immun} in Appendix \ref{sec:opinions}.}

\subsubsection{Did exposure lead to changes in immunization rate?} 
 
The last question to ask is whether, among those who knew their friends, neighbors, or relatives' immunization behavior, is there more immunizing behavior since June 2015 when there is more exposure to the campaign? That is, does the campaign appear change immunization behavior as reported by our survey respondents?

Panel B of Table \ref{tab:networks}  presents the results, where the dependent variable is whether the respondent knows of anyone among their friends, neighbors, or family who immunized a child. Specifically, we condition the sample when we look at knowledge among the network to those who knew whether the vaccination status of the children of members of their network (neighbors, friends, and family), so this effect is distinct from the effects reported above. Then we ask whether exposure to treatment is associated with greater incidence of vaccination take-up within their network. 

An important caveat to this table is that knowledge of the network members' children's vaccination status itself is affected by treatment (as discussed above). This is particularly true for the network of neighbors. If knowledge is itself affected by treatment, there can be at least two natural interpretations of these results. The first interpretation is that if the increase in reporting one's child's vaccination status to one's friend is unrelated to the actual status itself, then this regression picks up the treatment effect on vaccination itself. The second interpretation is that the treatment causes differential reporting of status to one's neighbors, so those who took up vaccination for their children may be more likely to speak about it. This is consistent with the results of \cite{banerjee2013diffusion}, where microfinance takers were nearly seven times more likely to pass on information about microcredit than those who were informed about it but chose not to adopt. To take another example, with more exposure, one may attend neighborhood health centers more and therefore observes more neighbors who take-up there. We cannot separate between the two interpretations, but we present the results nonetheless, advising that the estimates are to be interpreted with some caution. 

   Column 1 shows that when looking at the network of neighbors, an increased exposure by 15 tweets corresponds to a 12.5 percent  increase 
   in the number of vaccinations in the network (clustered $p = 0.071$, RI $p = 0.132$) relative to a mean of 0.356. When turning to the network of friends in column 2, we see an increased exposure by 15 tweets corresponds to a 16.0 percent  increase 
   in the number of vaccinations in the network (clustered $p = 0.001$, RI $p = 0.071$) relative to a mean of 0.353. Column 3 presents results looking at relatives. An increased exposure by 15 tweets corresponds to a 9.6 percent  increase 
   in the number of vaccinations in the network (clustered $p = 0.159$, RI  $p = 0.06$) relative to a mean of 0.314.
   
   Finally, Column 4 looks at own behavior, and our estimate  is not statistically different from zero. This is unsurprising as it is very rare that a respondent is exactly well-timed to have a child in the vaccination period during our experiment.

 \subsection{Discussion}
  
In this section we have shown that being randomly exposed to only 15 more campaign tweets on immunization leads to roughly a 10-20 percent  increase in awareness that an immunization campaign is going on, taking the awareness rate from a mean of 8-18 percent to 10-20 percent  overall. While each individual result is suggestive, taken together they paint a consistent picture: a moderate-intensity Twitter campaign seems to have changed people's beliefs and knowledge. It is worth noting that our campaign represents a mild exposure for a variety of reasons: people's feeds are flooded with many tweets so being exposed over multiple months to a dozen or so tweets on a specific topic may not be on anyone's radar. Moreover, this requires checking one's feed around the time that the tweet is sent out, which is a sparse event. Nonetheless, the mild increased awareness due to a light-touch exposure measurably led to large increases in the probability that respondents knew that vaccines were domestically produced. It also increased  offline conversations with network members so as to learn their (children's) vaccination statuses and possibly increased vaccination take-up behavior (or at least awareness of vaccination behavior in the network).

\

\section{Conclusion}\label{sec:conclusion}

We conducted a nationwide campaign, which consisted of randomized controlled trial on Twitter involving 37 celebrities and 9 organizations, to promote immunization and examine online diffusion and offline belief change.

We begin by studying what makes an effective campaign. First we decompose celebrity influence through reach and endorsement. We find that celebrity endorsement matters considerably. We then show that source citation has an adverse effect on diffusion.  Further, we show that there are decreasing returns to exposure: while the second exposure matters, individuals are no more likely to retweet after the third exposure to the campaign.

We then demonstrated that an online campaign can lead to changes in offline beliefs and knowledge, as measured by an separate phone survey. For instance, exposure to our campaign leads to both individuals knowing more about the details of immunization provision (e.g., know that they can keep halal while vaccinating their child due to domestic production). Further, individuals exposed to the campaign are much more likely to know the immunization seeking behavior and status of their friends, neighbors, and family, meaning that it likely spurred offline discussion. We also find suggestive evidence of higher reported immunization behavior changes by respondents' friends, relatives,  and neighbors. 

Our findings shed some light on what effective policy might look like. A simple rule of thumb is to recruit influential agents, like celebrities, to send messages without citing credible sources. With a fixed budget of messages, locating these messages such that there are multiple exposures, but not being wasteful,  would then generate the widest diffusion. 
All of this is not only  because mechanically influential agents are connected to more people, but also because subsequent individuals are considerably more likely to respond to the message when it is sent by influential agents such as celebrities (as evidenced by both likes and retweets).  
 Simply seeing the message 15 times over a four-month period can lead to sizeable shifts in offline knowledge and potentially even health-seeking behavior.

\

\bibliographystyle{ecta}
\bibliography{networks,nsf}

\newpage

\section*{Figures}


\begin{figure}[!h]
	\scalebox{0.75}{\begin{tikzpicture}[every text node part/.style={align=center}]
		\def \n {5}
		\def \radius {2cm}
		\def \margin {8}
		
		\node[draw, ellipse, minimum size=19pt] at (0,0) (v1){37 \\ Celebrities};
		
		\node[draw, ellipse, minimum size=19pt] at (-6,-5) (v2){Direct \\ Tweets};
		
		\node[draw, ellipse, minimum size=19pt] at (6,-5) (v3){Retweets};

		\node[draw, ellipse, minimum size=19pt] at (-9,-10) (v4){Without \\ Credibility \\ Boost};
		\node[draw, ellipse, minimum size=19pt] at (-3,-10) (v5){With \\ Credibility \\ Boost};
		
		\node[draw, ellipse, minimum size=19pt] at (3,-10) (v6){9 \\ Organizations};
		\node[draw, ellipse, minimum size=19pt] at (9,-10) (v7){1032 \\ Joes};

		\node[draw, ellipse, minimum size=19pt] at (1,-18) (v8){Without \\ Credibility \\ Boost};
		\node[draw, ellipse, minimum size=19pt] at (5,-18) (v9){With \\ Credibility \\ Boost};

		\node[draw, ellipse, minimum size=19pt] at (7,-15) (v10){Without \\ Credibility \\ Boost};
		\node[draw, ellipse, minimum size=19pt] at (11,-15) (v11){With \\ Credibility \\ Boost};

		\draw[line width = 0.3mm,->, >=latex] (v1) to (v2);
		\draw[line width = 0.3mm,->, >=latex] (v1) to (v3);
		\draw[line width = 0.3mm,->, >=latex] (v2) to (v4);
		\draw[line width = 0.3mm,->, >=latex] (v2) to (v5);
		\draw[line width = 0.3mm,->, >=latex] (v3) to (v6);
		\draw[line width = 0.3mm,->, >=latex] (v3) to (v7);

		\draw[line width = 0.3mm,->, >=latex] (v6) to (v8);
		\draw[line width = 0.3mm,->, >=latex] (v6) to (v9);
		\draw[line width = 0.3mm,->, >=latex] (v7) to (v10);
		\draw[line width = 0.3mm,->, >=latex] (v7) to (v11);
		
		\end{tikzpicture}}
	\caption{Experimental Design Schematic: Conditional on style, formal or casual, a half of the celebrity's tweets are assigned to direct and the other half are assigned to be retweeted. A subset of tweets (those that are not deemed sensitive by the government) are randomly assigned to be with or without crediblity boost. The retweets are either conducted by an organization, with probability 1/3, or a Joe with probability 2/3. 
		\label{fig:expt_design}}
\end{figure}
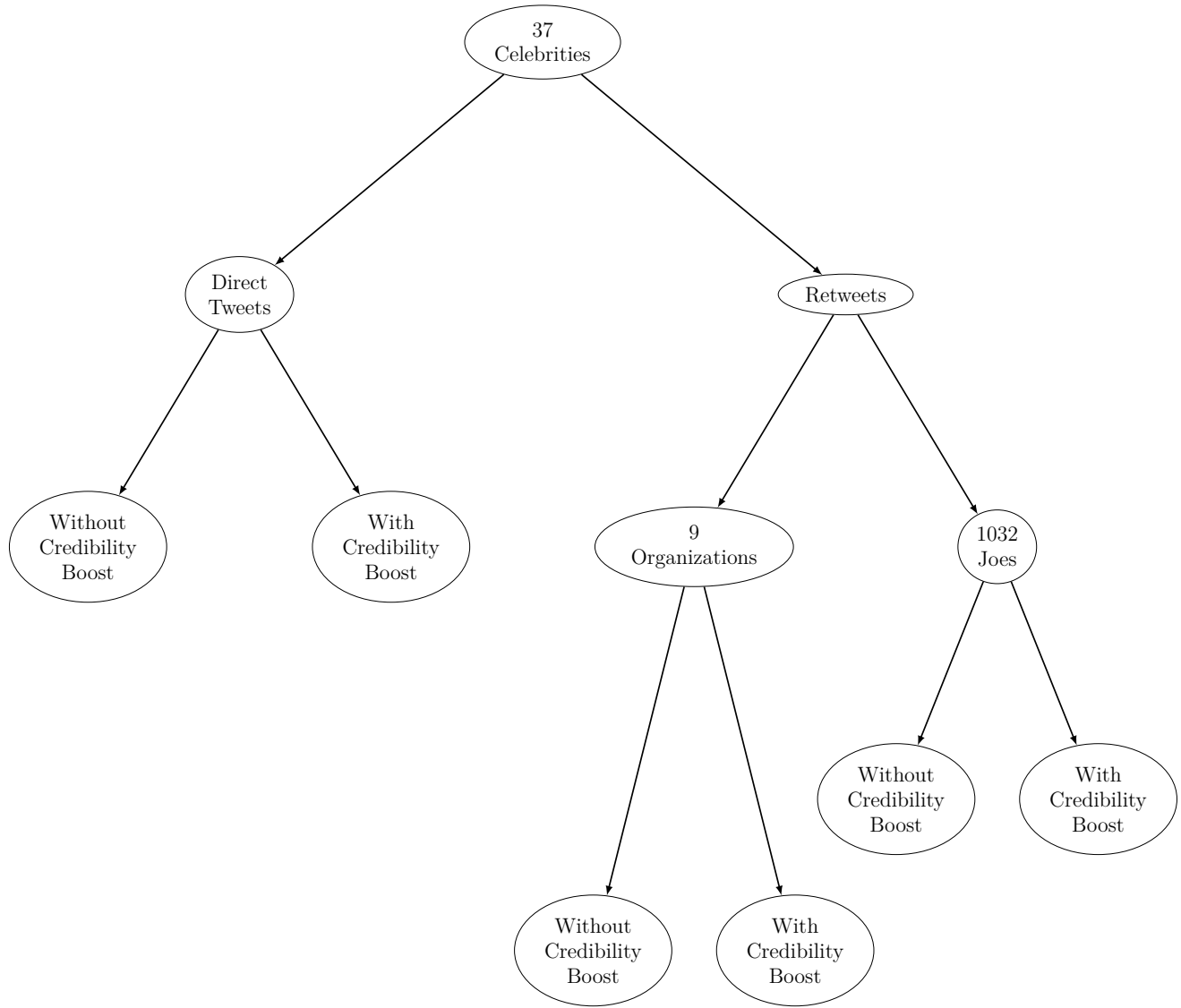

\clearpage


\clearpage

\clearpage


\clearpage

\clearpage

\newpage

\section*{Tables}


\begin{table}[!h]
\centering
\caption{User summary stats}\label{tab:user_summary}
\scalebox{1}{\begin{threeparttable}
\emph{Panel A: Online user summary}
{
\def\sym#1{\ifmmode^{#1}\else\(^{#1}\)\fi}
\begin{tabular*}{14cm}{@{\hskip\tabcolsep\extracolsep\fill}l*{1}{cccc}}
\hline\hline
                    &    &  &  mean&         obs\\
\hline
Followers of celebrities& &  &   262648&          37\\
Followers of organizations& & &     145300&           9\\
Followers of Joes/Janes   &        & & 574&         134\\
Followers of forced Joes/Janes&     &  &  502&         898\\
Followers of celeb followers&  &  &    1379&        1073\\
\hline\hline
\end{tabular*}
}

\emph{Panel B: Offline user summary}
{
\def\sym#1{\ifmmode^{#1}\else\(^{#1}\)\fi}
\begin{tabular*}{14cm}{@{\hskip\tabcolsep\extracolsep\fill}l*{1}{cccccc}}
\hline\hline
                    &   & & & &   Sample&      National Average (SUSENAS)\\
                    &   & & & & mean&        mean\\
\hline
Age                 &   & & & & 28.416&      29.959\\
Female              &   & & & & 0.504&       0.499\\
City (\emph{kota})              &   & & & & 0.610&       0.200\\
\hline\hline
\end{tabular*}
}

\emph{Panel C: Baseline beliefs}
{
\def\sym#1{\ifmmode^{#1}\else\(^{#1}\)\fi}
\begin{tabular*}{14cm}{@{\hskip\tabcolsep\extracolsep\fill}l*{1}{cc}}
\hline\hline
                    &        mean&         obs\\
\hline
Immunization is important&       0.988&         886\\
Immunization is safe&       0.944&         886\\
Immunization is beneficial&       0.983&         886\\
Breastfeeding can't replace immunization&       0.650&         886\\
Supplements can't replace immunization&       0.872&         886\\
Herbal supplements can't replace immunization&       0.832&         886\\
BCG is a basic vaccine&       0.452&         622\\
Hepatitis B is a basic vaccine&       0.291&         622\\
DPT is a basic vaccine&       0.399&         622\\
HIB is a basic vaccine&       0.084&         622\\
Polio is a basic vaccine&       0.712&         622\\
Measles is a basic vaccine&       0.611&         622\\
Immunization does not cause swelling&       0.561&         886\\
Immunization does not cause fever&       0.647&         886\\
Vaccines are domestically made&       0.561&         886\\
Vaccines are free of cost&       0.385&         886\\
\hline\hline
\end{tabular*}
}

\end{threeparttable}}
\end{table}

\clearpage


\begin{table}[!h]
	\centering
	\caption{Reach vs. Endorsement: Value of Celeb Endorsement through Involvement measured by $F_2$ likes/retweets}\label{tab:celeb_v_org_v_joe_F2}
	\scalebox{0.95}{\begin{threeparttable}
			\begin{tabular}{lcccccc} \hline
 & (1) & (2) & (3) & (4) & (5) & (6) \\
 & Poisson & Poisson & Poisson & Poisson & Poisson & Poisson \\
VARIABLES & \# Pooled & \# Pooled & \# Retweets & \# Retweets & \# Likes & \# Likes \\ \hline
 &  &  &  &  &  &  \\
Celeb writes and tweets & 0.544 & 0.788 & 0.518 & 0.931 & 0.664 & 1.109 \\
 & (0.166) & (0.505) & (0.166) & (0.584) & (0.482) & (0.687) \\
 & [0.00105] & [0.119] & [0.00175] & [0.111] & [0.168] & [0.107] \\
 &  &  &  &  &  &  \\
Observations & 1,997 & 911 & 1,997 & 911 & 1,997 & 911 \\
Joe/Jane writes mean & 0.0417 & 0.00915 & 0.0417 & 0.00686 & 0.00745 & 0.00229 \\
 Forced Joes/Janes only &  & \checkmark &  & \checkmark &  & \checkmark \\ \hline
\end{tabular}

			\begin{tablenotes}
				Notes: Standard errors (clustered at the original tweet level) are reported in parentheses. $p$-values are reported in brackets. Sample conditions on all tweets originated by Joes/Janes or celebrities. All regressions control for phase, celebrity fixed effects, content fixed effects, and the log number of followers of the $F_1$.
			\end{tablenotes}
	\end{threeparttable}}
\end{table}

\begin{table}[!h]
	\centering
	\caption{Value of Celeb Endorsement through Composition measured by $F_1$ likes/retweets}\label{tab:celeb_v_org_v_joe_F1}
	\scalebox{1}{\begin{threeparttable}
			\begin{tabular}{lccc} \hline
 & (1) & (2) & (3) \\
 & Poisson & Poisson & Poisson \\
VARIABLES & \# Pooled & \# Retweets & \# Likes \\ \hline
 &  &  &  \\
Celeb writes and tweets & 1.101 & 1.329 & 0.803 \\
 & (0.0840) & (0.0910) & (0.105) \\
 & [0] & [0] & [0] \\
 &  &  &  \\
Observations & 451 & 451 & 451 \\
 Joe/Jane writes and Celeb retweets mean & 2.058 & 1.045 & 1.013 \\ \hline
\end{tabular}

			\begin{tablenotes}
				Notes: Standard errors (clustered at the original tweet level) are reported in parentheses. $p$-values are reported in brackets. Sample conditions on all tweets originated by Joes/Janes or celebrities. All regressions control for phase, celebrity fixed effects, and content fixed effects.
			\end{tablenotes}
	\end{threeparttable}}
\end{table}

\clearpage

\begin{landscape}
	\begin{table}[!h]
		\centering
		\vspace*{2.5cm}
		\caption{Value of Endorsement through Source Citation measured by $F_1$ likes/retweets}\label{tab:credboost_F1}
		\scalebox{0.81}{\begin{threeparttable}
				\begin{tabular}{lcccccccccccc} \hline
 & (1) & (2) & (3) & (4) & (5) & (6) & (7) & (8) & (9) & (10) & (11) & (12) \\
 & Poisson & Poisson & Poisson & Poisson & Poisson & Poisson & Poisson & Poisson & Poisson & Poisson & Poisson & Poisson \\
VARIABLES & \# Pooled & \# Pooled & \# Pooled & \# Pooled & \# Retweets & \# Retweets & \# Retweets & \# Retweets & \# Likes & \# Likes & \# Likes & \# Likes \\ \hline
 &  &  &  &  &  &  &  &  &  &  &  &  \\
Source cited & -0.306 & -0.553 & -0.235 & -0.0365 & -0.318 & -0.694 & -0.347 & 0.0946 & -0.277 & -0.261 & 0.0104 & -0.239 \\
 & (0.157) & (0.248) & (0.109) & (0.207) & (0.161) & (0.297) & (0.113) & (0.186) & (0.183) & (0.236) & (0.248) & (0.297) \\
 & [0.0513] & [0.0260] & [0.0319] & [0.860] & [0.0478] & [0.0195] & [0.00222] & [0.612] & [0.130] & [0.269] & [0.967] & [0.421] \\
 &  &  &  &  &  &  &  &  &  &  &  &  \\
Observations & 492 & 170 & 131 & 191 & 492 & 170 & 131 & 191 & 492 & 170 & 131 & 191 \\
Depvar Mean & 3.644 & 2.635 & 7.305 & 2.031 & 3.644 & 2.635 & 7.305 & 2.031 & 3.644 & 2.635 & 7.305 & 2.031 \\
Celeb RT Joe/Jane &  & \checkmark &  &  &  & \checkmark &  &  &  & \checkmark &  &  \\
Celeb RT Org &  &  &  & \checkmark &  &  &  & \checkmark &  &  &  & \checkmark \\
 Celeb Direct &  &  & \checkmark &  &  &  & \checkmark &  &  &  & \checkmark &  \\ \hline
\end{tabular}

				\begin{tablenotes}
					Notes: Standard errors (clustered at the celebrity/organization level) are reported in parentheses. $p$-values are reported in brackets.   All regressions control for phase, celebrity fixed effects, content fixed effects, and condition on non-sensitive tweets.
				\end{tablenotes}
		\end{threeparttable}}
	\end{table}
\end{landscape}

\clearpage

\begin{table}[!h]
	\centering
	\scalebox{1}{\begin{threeparttable}
			\caption{Shape of response function: Value of exposure}\label{tab:panel_daylevel_exposure}
			\begin{tabular}{lc} \hline
 & (1) \\
 & Poisson \\
VARIABLES & Retweet \\ \hline
 &  \\
Exposure to 2 tweets today & 0.740 \\
 & (0.293) \\
 & [0.0114] \\
Exposure to 3 tweets today & 0.948 \\
 & (0.549) \\
 & [0.0839] \\
Exposure to 4-7 tweets today & 0.703 \\
 & (0.890) \\
 & [0.430] \\
 &  \\
Observations & 15,263 \\
Potential tweet exposure control & \checkmark \\
User FE & \checkmark \\
Phase control & \checkmark \\
Date FE & \checkmark \\
 Depvar Mean & 0.0300 \\ \hline
\end{tabular}

			\begin{tablenotes}
				Notes: Standard errors (clustered at user level) are reported in parentheses. $p$-values are reported in brackets. 
			\end{tablenotes}
	\end{threeparttable}}
\end{table}

\clearpage


\begin{table}[!h]
\caption{Did people offline hear about the campaign?}\label{tab:heard_campaign}
\centering
\scalebox{1}{\begin{threeparttable}
\begin{tabular}{lccc} \hline
 & (1) & (2) & (3) \\
 & Logit & Logit & Poisson \\
 & Heard of & Heard of & \# of times \\
VARIABLES & $\#Ayoimunisasi$ & immunization from Twitter & heard from Twitter \\ \hline
 &  &  &  \\
Std. Exposure to tweets & 0.197 & 0.108 & 0.106 \\
 & (0.0980) & (0.0666) & (0.0529) \\
 & [0.0443] & [0.106] & [0.0444] \\
 & \{.107\} & \{.046\} & \{.127\} \\
 &  &  &  \\
Observations & 2,164 & 2,404 & 2,441 \\
Potential exposure control & \checkmark & \checkmark & \checkmark \\
Double Post-LASSO & \checkmark & \checkmark & \checkmark \\
 Depvar Mean & 0.0776 & 0.181 & 0.322 \\ \hline
\end{tabular}

\begin{tablenotes}
Notes: Standard errors (clustered at the combination of celebs followed level) are reported in parentheses. Clustered $p$-values are reported in brackets. Randomization inference (RI) $p$-values are reported in braces. Demographic controls include age, sex, province, dummy for urban area and dummy for having children. One standard deviation of exposure is 14.96 tweets.
\end{tablenotes}
\end{threeparttable}}
\end{table}

\clearpage

\begin{table}[!h]
	\caption{Did people offline increase knowledge?}\label{tab:knowledge_immun}
	\centering
	\scalebox{1}{\begin{threeparttable} 
			\begin{tabular}{lcccccc} \hline
 & (1) & (2) & (3) &  (4)   \\
 & Logit &  Logit & Logit & Logit   \\
VARIABLES &    Domestic &   Substitutes &  Side-effects & Free \\ \hline
 &  &    &  &  \\
Std. Exposure to tweets   & 0.120 & -0.0391 & 0.0305 & 0.0549 \\
   & (0.0591) & (0.0589) & (0.0624) & (0.0687) \\
   & [0.0424] & [0.506] & [0.625] & [0.424] \\
    & \{.028\} & \{.891\} & \{.751\} & \{.629\} \\
 &     &  &  &  &  \\
Observations   & 2,434 & 2,440 & 2,440 & 2,440 \\
Potential exposure control    & \checkmark & \checkmark & \checkmark & \checkmark \\
Double Post-LASSO    & \checkmark & \checkmark & \checkmark & \checkmark \\
 Depvar Mean   & 0.576 & 0.527 & 0.486 & 0.680 \\ \hline
\end{tabular}

			\begin{tablenotes}
				Notes: Standard errors (clustered at the combination of celebs followed level) are reported in parentheses. Clustered $p$-values are reported in brackets. Randomization inference (RI) $p$-values are reported in braces. Demographic controls include age, sex, province, dummy for urban area and dummy for having children. One standard deviation of exposure is 14.96 tweets.  
			\end{tablenotes}
	\end{threeparttable}}
\end{table}

\clearpage

\begin{table}[!h]
	\caption{Networks and behavior}\label{tab:networks}
	\centering
	\scalebox{1}{\begin{threeparttable}
		
		\emph{Panel A: Knowledge of neighbor, friend, and relative network members' immunization behavior}
		\begin{tabular}{lccccc} \hline
 & (1) & (2) & (3)  \\
  &  Logit  & Logit & Logit \\
VARIABLES &   Neighbor & Friend & Relative \\ \hline
 &   &  &  \\
Std. Exposure to tweets   & 0.231 & 0.0156 & 0.214 \\
  & (0.0814) & (0.0826) & (0.132) \\
  & [0.00449] & [0.850] & [0.105] \\
   & \{.088\} & \{.778\} & \{.462\} \\
 &   &  &  \\
Observations   & 1,642 & 1,626 & 1,564 \\
Potential exposure control   & \checkmark & \checkmark & \checkmark \\
Double Post-LASSO   & \checkmark & \checkmark & \checkmark \\
 Depvar Mean  & 0.775 & 0.813 & 0.923 \\ \hline
\end{tabular}

			\emph{Panel B: Immunization behavior of neighbors, friends, and relatives network members as well as self}
						\begin{tabular}{lccc | c } \hline
 & (1) & (2) & (3) & (4)   \\
  & Logit & Logit & Logit & Logit \\
VARIABLES &   Neighbor & Friend & Relative &   Own \\ \hline
 &    &  &  &  \\
Std. Exposure to tweets  & 0.194 & 0.246 & 0.140 & -0.0840 \\
  & (0.107) & (0.0955) & (0.0997) & (0.0886) \\
 & [0.0707] & [0.00994] & [0.159] & [0.343] \\
 & \{.132\} & \{.071\} & \{.06\} & \{.66\} \\
   &  &  &  &  \\
Observations   & 682 & 682 & 682 & 634 \\
Potential exposure control   & \checkmark & \checkmark & \checkmark & \checkmark \\
Double post-LASSO  & \checkmark & \checkmark & \checkmark & \checkmark \\
 Depvar Mean   & 0.356 & 0.353 & 0.314 & 0.486 \\ \hline
\end{tabular}

			\begin{tablenotes}
				Notes: In both panels, standard errors (clustered at the combination of celebs followed level) are reported in parentheses. Clustered $p$-values are reported in brackets. Randomization inference (RI) $p$-values are reported in braces. Demographic controls include age, sex, province, dummy for urban area and dummy for having children. One standard deviation of exposure is 14.96 tweets. In Panel A, the sample is  restricted to respondents who know friends/relatives/neighbors with at least one child (ages 0-5) respectively. In Panel B, the sample when looking at network members' behaviors (columns 1-3) is restricted to respondents who know the behavior of their network. When looking at own behavior in column 4, sample restricted to respondents with children younger than age 2.
			\end{tablenotes}
	\end{threeparttable}}
\end{table}

\clearpage

\appendix
\newpage

\clearpage

\section{Model\label{sec:Model}}

\subsection{Overview}

We study the decision by individuals on Twitter to pass on information
to their followers by ``retweeting'' it. Before proceding to our
empirical analysis, we begin by discussing a simple framework to think
through how individuals make the decision to pass on information. The framework is standard, developed in \cite*{chandrasekhar2018signaling} and  also previously applied in \cite*{banerjee2018less}. 

In our framework, individuals pass on information for two reasons.
First, individuals may care that others are informed about a topic.
Second, as retweeting is intrinsicially a social activity, individuals
can be motivated by how they are viewed by their followers. In this
case, individuals may choose to retweet certain topics as a function
of how the act of sharing the information changes how they are perceived
by others. For example, individuals on Twitter may be trying to gather
more followers, and it is plausible that people are more likely to
keep following someone whom they believe is sharing high-quality information.

This second observation---that people may share information
with a view to it affects how others perceive them---turns
out to have subtle ramifications for how we think about a dissemination
strategy. Whether information is more likely to spread more widely
if originated by a celebrity or an ordinary Joe, whether messages
cite credible sources or simply consist of assertions,
whether the messaging involves multiple exposures, and even whether
the topic consists of more well-known ideas or dispelling persistent
myths, turn out to be ambiguous questions once we include the fact
that these features of messages change the degree to which sharing
the message provides information in equilibrium about the likely quality
of the person deciding whether to share it.

In particular, the standard intuition is that that more and credible
information is simply better, and hence more likely to be retweeted.
This comes from a standard model in which individuals only base their
decisions to pass on information based on the first factor, namely
the quality of that information. In this case, if a message has more
credibility, has a verified source, has a history of multiple exposure,
or has higher marginal value (in the sense that it dispels otherwise
wrong beliefs), the more retweeting should happen. This generates
an intuition that, for instance, sourced tweets or celebrity tweets
should be retweeted more.

However, when we consider the fact that retweeting has a social component\textemdash that
individuals certainly care about how they are perceived and that is
likely a key component of their motivation to retweet\textemdash we
see that these conclusions change. Assume that an individual $F$
follows an originator of a tweet, $o$. Suppose that $F$ is more
willing to pass on information if he is more certain about the state
of the world. Also assume that $F$ can be one of two private types:
a high type (greater ability or social consciousness for the sake
of discussion) and a low type. Individuals desire to be perceived
of as a high type by their followers, so part of the motivation to
retweet is for this social perception payoff. It is commonly known
that high types are better able to assess the state of the world rather
than low types (i.e., imagine that in addition to the tweet, individual
$F$ gets a private signal as to the state of the world, and the high
types' signal is more informative). When $F$ sees a tweet by $o$,
he needs to glean the state of the world using both the tweet and
his own private signal, and decide whether or not to retweet. 

To illustrate ideas, let us compare the case where $o$'s tweet contains
no source versus cites a credible source about the topic. Inclusion
of a source has multiple effects. First, the source citation should
make the state of the world even more evident. This should encourage
retweeting through increasing certainty. Second, and more subtly,
if social perception is important enough, source citation can have
a discouraging effect on retweeting. Specifically, if a source makes
it very clear what is true, then there is no room for signaling remaining:
high types are no better able to assess things than low types and
therefore ability does not really matter. We show below that which
effect dominates on net---the increased direct effect
of the source on quality, or the fact that the source decreases the
ability of $F$ to use the tweet to signal quality---turns
out to be ambiguous.

To show this more formally, we adapt the endogenous communication
model developed by \cite*{chandrasekhar2018signaling} to our context
of retweeting on Twitter (see also \cite{banerjee2018less} for another such prior  application of this model). Such image concerns have also been looked both theoretically and empirically in both \cite*{bursztyn2015does,bursztyn2017cool} who  study whether peer perceptions inhibit the seeking of
 education. We look at individuals who have payoffs
from passing on information and who are concerned with social perception
as well the direct value of the information they pass. We show how
that sourcing, originator identity, exposure, and content all can
have ambiguous effects on the amount of retweeting, and explore when
we might expect which policies to work well.

It is important to note that we are not claiming of course that these
are the only motives for retweeting. After all there can be more mundane
motivations: it is just more fun to retweet anything by a celebrity,
or it is just frivolous to retweet anything by a celebrity, for instance.
But without hardcoding anything else into the model, in the simplest
interpretation of dynamics on Twitter, we can demonstrate and motivate
why ultimately the questions we study are ultimately empirical issues.

\subsection{Setup}

\subsubsection{Environment}

The state of the world is given by $\eta\in\left\{ 0,1\right\} $,
with each state equally likely. There is an originator $o$ (she)
who writes an initial message about the idea with probability $q\in(0,1]$,
which is received by her follower $F$ (he). With probability $1-q$
nothing happens. The message is a binary signal about the state of
the world, which is accurate with probability $\alpha$, i.e. 
\[
m=\begin{cases}
\eta & \text{w.p. }\alpha\geq\frac{1}{2}\\
1-\eta & \text{o.w.}
\end{cases}.
\]
The message may or may not cite a source, designated by $z\in\left\{ S,NS\right\} $
respectively. We allow the quality of the signal to depend on source,
so $\alpha=\alpha_{z}$, discussed below.

Further, there are two types of originators: ordinary Janes/Joes and celebrities,
given by $o\in\left\{ J,C\right\} $ respectively. We allow the quality
of the signal to depend on originator, so $\alpha=\alpha_{o}$, discussed
below.

Finally, followers come in two varieties: $\theta\in\left\{ H,L\right\} $
represents $F$'s privately known type, and one's type is drawn with
equal odds. High types have better private information about the state
of the world. This can represent ability in a loose way such as intelligence,
social accumen, taste-making ability, or any trait which allows $F$
to better discern the state of the world if he is of type $H$ rather
than $L$. We model this by supposing that $F$ draws an auxiliary
signal, $x$, with $x=\eta$ with probability $\pi_{\theta}$ and
$x=1-\eta$ with probability $1-\pi_{\theta}$. We assume $\pi_{H}\geq\pi_{L}$
which reflects that $H$-types can better discern whether the idea
is valuable. As discussed below, it is socially desirable to be perceived
as $\theta=H$. 

\

This environment captures our basic experimental setting. We randomly
vary originator $o\in\left\{ J,C\right\} $ and whether the message
is sourced, $z\in\left\{ S,NS\right\} $. We also vary whether the
topic concerns a myth, which is modeled by $F$ having worse beliefs
about the true state of the world $\pi_{\theta}^{myth}\leq\pi_{\theta}^{non-myth}$,
and varying the number of exposures which is simply modeled by the
number of prior signals that $F$ has received.

\subsubsection{Bayesian Updating}

$F$ is assumed to be Bayesian. Let $\alpha=\alpha_{o,z}$ be the
quality of the signal depending on originator and source. Therefore
given message $m$ and private signal $x$, we can compute the likelihood
ratio that $F$ believes the state of the world being good versus
bad as
\begin{align*}
LR\left(\eta\vert m,x;o,z,\theta\right) & =\frac{\Pr\left(\eta=1\vert m,x\right)}{\Pr\left(\eta=0\vert m,x\right)}=\frac{\Pr\left(m,x\vert\eta=1\right)}{\Pr\left(m,x\vert\eta=0\right)}\\
& =\left(\frac{\alpha_{o,z}}{1-\alpha_{o,z}}\right)^{m}\left(\frac{1-\alpha_{o,z}}{\alpha_{o,z}}\right)^{1-m}\left(\frac{\pi_{\theta}}{1-\pi_{\theta}}\right)^{x}\left(\frac{1-\pi_{\theta}}{\pi_{\theta}}\right)^{1-x}.
\end{align*}
Note that as $\alpha$ or $\pi$ tend to 1 or $\frac{1}{2}$, the
likelihood ratio tends to $+\infty$ (the signal reveals the state)
or 1 (the signal has no content), respectively.

\subsubsection{Payoffs}

The utility of $F$ depends on two components. The first is the instrumental
payoff: it is a payoff from retweeting when the state of the world
is more clear: that is when $LR\left(\eta\right)$ is more extreme.
Thus we assume that the instrumental payoff when you do not retweet,
i.e., when $r=0$, is 0 and when you do retweet, i.e., $r=1$, is
$\varphi\left(LR\left(\eta\vert m,x;o,z,\theta\right)\right)$ for
some smooth increasing in distance function from $1$, $\varphi\left(\cdot\right)$.
What this captures is that there is more instrumental value in passing
on a message the greater certainty in the state of the world. For
instance if we set 
\[
\varphi\left(x\right)=f\left(\left|\frac{x}{1+x}-1\right|\right)
\]
for a smooth increasing function $f\left(\cdot\right)$ on $\left[0,\frac{1}{2}\right]$,
the instrumental value is a monotone function in the probability the
state of the world is high, but other functions $\varphi$ will also
work.\footnote{To see this, note that
	\[
	\varphi\left(LR\left(\eta\vert m,x;o,z,\theta\right)\right)=f\left(\left|\frac{LR}{1+LR}-1\right|\right)=f\left(\left|\Pr\left(\eta=1\vert m,x;o,z,\theta\right)-\frac{1}{2}\right|\right)
	\]
	which is just a smooth function of distance from pure uncertainty
	of a belief of $\frac{1}{2}$.} Further, due to taste or cost heterogeneity, there is a shock $\epsilon$
to the instrumental payoff of retweeting, where $\epsilon$ is a mean-zero
random variable drawn from a continuous CDF with full support, such
as the logit CDF $\Lambda\left(\cdot\right)$. Altogether, the instrumental
payoff $V^{r}$ is given by
\[
V^{r}=\begin{cases}
\varphi\left(LR\left(\eta\vert m,x;o,z,\theta\right)\right)-\epsilon & \text{if }r=1\\
0 & \text{if }r=0.
\end{cases}
\]

The second is the social perception payoff. Specifically $F$ is concerned
with the posterior that his followers have about his type given his
decision to retweet: $\psi\left(\Pr\left(\theta=H\vert r\right)\right)$
where $\psi\left(\cdot\right)$ is a monotonically increasing function.
The perception in equilibrium is simply a function of the retweet
decision itself. The idea here is that someone who is more able is
more likely to be able to discern valuable topics and therefore the
equilibrium decision to retweet itself has a signaling component.
\footnote{For simplicity we abstract from $F$'s followers interpretation of
	$m$ and their own subsequent private signals. The reason is that
	we can demonstrate interesting non-monotonicities in retweeting behavior
	as a function of message quality without such additions, which would
	only serve to complicate matters.}

$F$'s total utility is given by
\[
U\left(r\vert m,x\right)=\underbrace{V^{r}}_{\text{instrumental}}+\underbrace{\lambda\psi\left(\Pr\left(\theta=H\vert r\right)\right)}_{\text{perception}}
\]
where $\lambda\geq0$ is a parameter that tunes the strength of the
perception payoff. \footnote{While $\lambda$ could be absorbed into $\psi\left(\cdot\right)$,
	it is useful for exposition to keep it separate.}

Correspondently the marginal utility of choosing $r=1$ versus $r=0$
is given by
\[
MU\left(r\vert m,x\right)=\underbrace{\varphi\left(LR\left(\eta\vert m,x;o,z,\theta\right)\right)-\epsilon}_{\text{change in instrumental}}+\underbrace{\lambda\Delta_{r}\psi\left(\Pr\left(\theta=H\vert r\right)\right)}_{\text{change in perception}}.
\]
Let $Q_{H}\left(\cdot\right)$ be the CDF of $\varphi\left(LR\left(\eta\vert m,x;o,z,H\right)\right)-\epsilon$
and $Q_{L}\left(\cdot\right)$ be the CDF of $\varphi\left(LR\left(\eta\vert m,x;o,z,L\right)\right)-\epsilon$.\footnote{This holds fixed $o$ and $z$.}
It immediately follows that $Q_{H}\succ_{\text{FOSD}}Q_{L}$. This
can be seen by inspection, where the likelihood ratio under type $H$
first order stochastically dominates that of type $L$ when $\eta=1$
and the inverse of the ratio first order stochastically dominates
when $\eta=0$. It will be useful below to denote by $G_{\theta}$
the complementary CDF, $G_{\theta}:=1-Q_{\theta}$, i.e., $G_{\theta}(v)$
is the fraction of types $\theta$ with a (net-of-costs) instrumental
value of passing greater than or equal to $v$. 

\subsection{Analysis}

$F$ decides to retweet if and only if $MU\left(r\vert m,x\right)\geq0$.
This decision trades off two components. On the one hand is the relative
instrumental benefit (or cost) of pasing on the message, which is
an increasing function of the likelihood that the state of the world
$\eta=1$, and is given by $\varphi\left(LR\left(m,x\vert o,z,\theta\right)\right)$.
On the other hand, retweeting itself changes the perception of $F$
by his followers, given by$\Delta_{r}\psi\left(\Pr\left(\theta=H\vert r\right)\right)$,
and so the (equilibrium) relative gain/loss of reputation must be
taken into account.

The model is formally characterized in Proposition 1 of Chandrasekhar,
Golub, and Yang (2018), and we refer the interested reader to that
paper for proofs. Chandrasekhar, Golub, and Yang show that under the
above assumptions, an equilibrium exists, and will be in cutoff strategies
where $F$ chooses to retweet if and only if $\varphi\left(LR\left(\eta\vert m,x;o,z,\theta\right)\right)-\epsilon\geq v$
for some $v$. An equilibrium is characterized by a cutoff $\underline{v}<0$,
which is used by all $F$s irrespective of type $\theta$, where it
is the solution to
\[
\underline{v}=\lambda\psi\left(\Pr\left(\theta=H\vert r=0\right)\right)-\lambda\psi\left(\Pr\left(\theta=H\vert r=1\right)\right).
\]
Here the equilibrium posteriors are determined by:
\[
\frac{\Pr\left(\theta=H\vert r=0\right)}{1-\Pr\left(\theta=H\vert r=0\right)}=\frac{1-qG_{H}\left(v\right)}{1-qG_{L}\left(v\right)}\text{ and }\frac{\Pr\left(\theta=H\vert r=1\right)}{1-\Pr\left(\theta=H\vert r=1\right)}=\frac{G_{H}\left(v\right)}{G_{L}\left(v\right)}.
\]

The intuition for the equilibrium is as follows. First, note that
$F$'s type does not matter for the decision he makes conditional
on the draw $v$. That is, while $\theta$ affects the distribution
of the instrumental value, once $F$ knows his instrumental value,
he is trading off that against the change in reputation due to his
behavior. Therefore the cutoff (in utility space) will not depend
on $\theta$'s type.

At the cutoff $\underline{v}$ in equilibrium the marginal benefit
of retweeting (which is a way to gain reputation by being viewed as
more likely to be a high type) must be equal to the marginal cost
of retweeting (which in this case is the instrumental benefit of passing
the information relative to the stochastic cost). The reason $\underline{v}<0$
is because here retweeting is a signal of being the high type, and
therefore some low types will opt into retweeting despite having a
negative net instrumental cost.

Holding fixed $o,z$ as we have been doing above, we can compute the
retweeting share in equilibrium:
\[
\frac{1}{2}G_{H}\left(\underline{v}\right)+\frac{1}{2}G_{L}\left(\underline{v}\right).
\]

We can also look at several contrasting situations. In the first,
assume that $\lambda=0$ with the same setup as above, so there is
no interest in social concerns. Then only positive instrumental values
are retweeted, so the share retweeting is given by
\[
\frac{1}{2}G_{H}\left(0\right)+\frac{1}{2}G_{L}\left(0\right).
\]
Clearly the retweeting share is lower than when there is also a signaling
motive, which featured an equilibrium cutoff $\underline{v}<0$.

A second contrasting situation is one in which while individuals would
potentially care about signaling, neither party is better at discerning
the state of the world. That is, $\widehat{G}_{H}=\widehat{G}_{L}=:\widehat{G}$.
In this case the share retweeted again is only determined by positive
instrumental values and therefore is given by
\[
\widehat{G}\left(0\right).
\]
Whether $\widehat{G}\left(0\right)\lesseqgtr\frac{1}{2}G_{H}\left(\underline{v}\right)+\frac{1}{2}G_{L}\left(\underline{v}\right)$
depends on how $\widehat{G}$ compares to $G_{H}$ and $G_{L}$.

\

A subtle feature of the model is the fact that the retweet share is
not necessarily monotonically increasing in the quality of the message,
$\alpha$. Intuitively, there are two effects of increasing $\alpha$
of retweeting. First, as $\alpha$ increases, the message becomes
more informative. This increases the instrumental value of retweeting,
and hence retweeting increases with $\alpha$. Second, as $\alpha$
increases, the $m$ signal becomes more informative relative to the
private $x$ signal. This makes the act of retweeting more about $m$
than $x$, and hence lowers the signaling value of retweeting. Indeed,
in the limit where $\alpha=1$, there is no signaling value whatsoever.
Thus, the signaling effect leads to a reduction in the amount of retweeting
as $\alpha$ increases. Which effect dominates depends on parameters,
and as we show now, in fact the effect of $\alpha$ on retweeting
can be non-monotonic under some configurations of parameters. 

\begin{figure}
	\centering{}\includegraphics[scale=0.85]{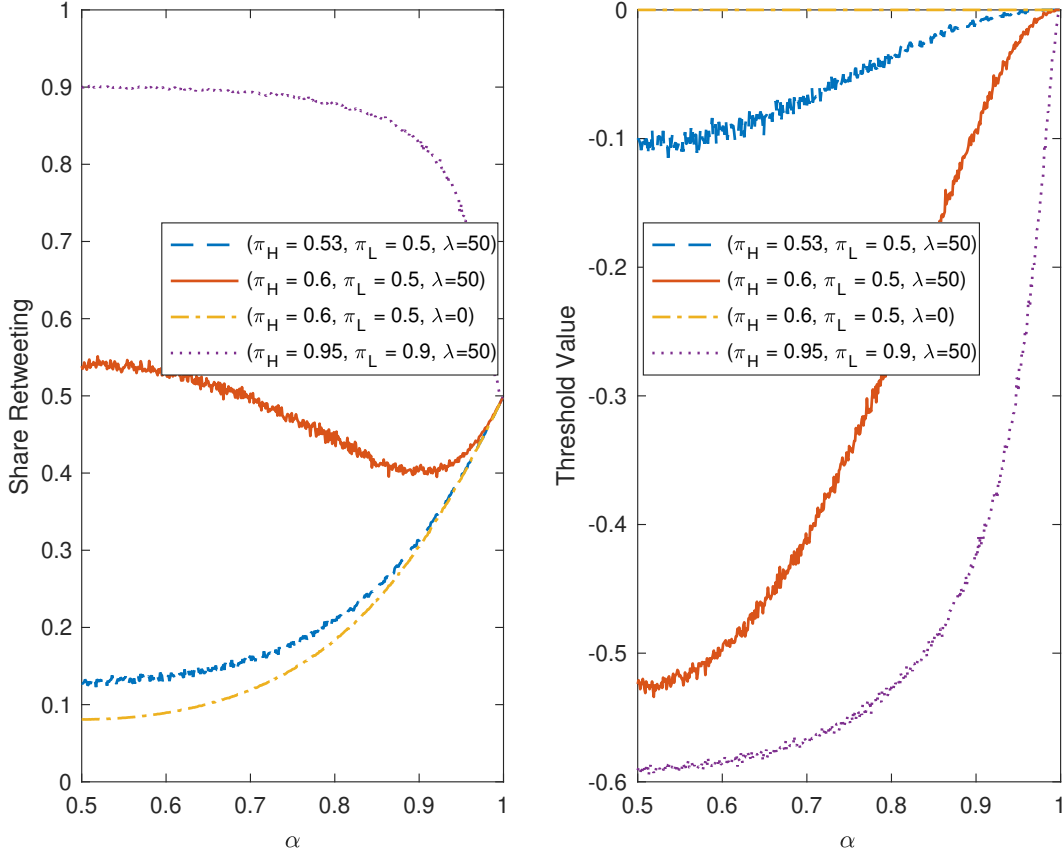}\caption{Retweet share for various combinations of $\left(\pi_{H},\pi_{L},\lambda\right)$.\label{fig:retweet-sim}}
\end{figure}

Figure \ref{fig:retweet-sim} presents simulation results to further
illustrate these intuitions. First consider the case when there is
no reputation considerations ($\lambda=0$). In this case, as the
message's quality increases, the share retweeting must increase clearly
because the value of information on average increases.

Next let us consider the case where neither $H$ nor $L$ are particularly
able types, with $\pi_{H}=0.53$ and $\pi_{L}=0.5$. In this case,
there is limited scope for signaling because the priors are quite
poor: both types heavily lean on the message's signal $m$ rather
than their personal signals $x$. As such, like in the case with $\lambda=0$,
the quality of the message increases the share to retweet.

In contrast, consider the case where both types are expert, but $H$-types
are somewhat better ($\pi_{H}=0.95$, $\pi_{L}=0.9$). In this case,
with low $\alpha$, since the predominant component of instrumental
value comes from type to begin with, and because high types are much
more likely to receive correct signals than low types but both have
typically good signals about the state of the world (so $m$ and $x$
will agree), many more $L$ types will also find it worthwhile to
essentially ``pool'' with $H$ types despite negative instrumental
values due to reputation concerns. This leads to a monotonic decline
in the retweet rate as $\alpha$ increases, since there is increasing
reliance on the $m$-signal. What this means in practice is that it
is possible to improve the quality of the message and yet decline
the overall share of retweeting, contrary to the naive intuition without
a social perception payoff component.

The final case we show is the intermediate one, with $\pi_{H}=0.65$
and $\pi_{L}=0.5$. The signaling effect at this parameter level dominates
initially, and hence increasing $\alpha$ initially decreases retweeting,
but then eventually is dwarfed by the instrumental effect as the $m$-signal
is considerably better than the gap in quality for the $x$-signal
across types. 

The fact that the relationship between $\alpha$ and retweeting is
non-monotonic means that it is possible that mild increases in informativeness
can reduce retweeting whereas dramatic increases in informativeness
can increase it.

\subsection{Application to Experiment}

In what follows, we use the above framework consider possible implications
of our experimental variations, i.e., (1) whether the originator is
a celebrity or a Jane/Joe, (2) whether the tweet has a source or not, (3)
whether the tweet concerns a myth or not, and (4) whether $F$ has
received prior exposures. We ignore indexes that are unnecessary in
each case.

\subsubsection{Celebrity versus Jane/Joe}

Celebrities and Joes can vary in the quality of their messaging. As
such, we consider $\alpha_{C}$ versus $\alpha_{J}$. Ex ante it may
be possible for these to have any relationship, though we might think
that celebrities tend to generate higher-quality signals. This could
be, for instance, because celebrities' messages reach many more individuals
and therefore they need to be more cautious in their messaging, or
could be because they have better access to information in general.

Assuming $\alpha_{C}\geq\alpha_{J}$ and since $\eta=1$ for an experimental
topic (since all our messages are sent about true beneficial effects
of immunization), 
\[
{\rm E}_{m,x}\left[\varphi\left(LR\left(\eta\vert m,x;C,\theta\right)\right)\right]\geq{\rm E}_{m,x}\left[\varphi\left(LR\left(\eta\vert m,x;J,\theta\right)\right)\right]
\]
and therefore the distribution of instrumental payoffs $Q_{C,\theta}\succ Q_{J,\theta}$
for each $\theta$. Note that this depends both on the originator
and the type of the individual.

To see the effect, consider the case when $\alpha_{C}\rightarrow1$.
In this case, following the intuition discussed above, $Q_{C,H}\rightarrow Q_{C,L}$
and let $\widehat{Q}_{C}\left(\cdot\right)$ be the resulting CDF
of the instrumental value, so there is nothing to signal at all. Thus
$\underline{v}^{C}=0$ and so anyone with any positive instrumental
value immediately retweets. In contrast, with Joes, as above there
is some negative $\underline{v}^{J}<0$ that sets the equilibrium.

Consequently, the retweeting share is given by
\begin{itemize}
	\item $\widehat{G}_{C}\left(0\right)$ under Celebrity origination and
	\item $\frac{1}{2}G_{J,H}\left(\underline{v}^{J}\right)+\frac{1}{2}G_{J,L}\left(\underline{v}^{J}\right)$
	under Joe origination.
\end{itemize}
Notice that it is not clear which dominates. On the one hand, since
$\eta=1$ is essentially revealed as $\alpha_{C}\rightarrow1$, $\widehat{G}_{C}$
has a higher mean than $G_{J,\theta}$ for either $\theta$. On the
other hand, the cutoff $\underline{v}^{J}$ can be considerably below
0 making the point of evaluating the $G_{J,\theta}$ CDFs at a lower
point. This is because the likelihood ratio distribution of knowing
that we are in a ``good'' world is not the same under celebrities
(where it is substantially more likely) and Joes/Janes (where it is less
likely, but there is a signaling effect reason to retweet).
\begin{rem}
	The total endorsement effect we identify in the experiment can be
	thought of being comprised of (a) a shift in instrumental value and
	(b) a shift in the threshold to retweet due to the signaling effect.
	To see this 
	\begin{align*}
	\frac{1}{2}\left[\widehat{G}_{C}\left(0\right)-G_{J,H}\left(\underline{v}^{J}\right)\right]+\frac{1}{2}\left[\widehat{G}_{C}\left(0\right)-G_{J,L}\left(\underline{v}^{J}\right)\right] & =\frac{1}{2}\left[\widehat{G}_{C}\left(0\right)-G_{J,H}\left(0\right)\right]+\frac{1}{2}\left[G_{J,z,H}\left(0\right)-G_{J,H}\left(\underline{v}^{J}\right)\right]\\
	& +\frac{1}{2}\left[\widehat{G}_{C}\left(0\right)-G_{J,L}\left(0\right)\right]+\frac{1}{2}\left[G_{J,z,L}\left(0\right)-G_{J,L}\left(\underline{v}^{J}\right)\right].
	\end{align*}
	In this expression, the $\widehat{G}_{C}\left(0\right)-G_{J,\theta}\left(0\right)$
	term measures how for a given cutoff of 0, the amount of retweets
	increases when a message is originated by the celebrity, and the the
	$G_{J,\theta}\left(0\right)-G_{J,\theta}\left(\underline{v}^{J}\right)$
	term measures the change in the share of retweets when we move the
	cutoff to the left due to the signaling effect, holding the distribution
	fixed. When the signaling impetus is dominant, this second term can
	overtake the prior term, making even a celebrity originator generate
	a lower volume of retweets.
\end{rem}

\subsubsection{Sourcing}

In this case, holding originator fixed, we study the effect of adding
a source. The analysis is identical to the case with celebrities.
Ex-ante it seems reasonable to model sourcing as having direct positive
effect on the likelihood of the signal being true: $\alpha_{S}\geq\alpha_{NS}$.
Consequently 
\[
{\rm E}_{m,x}\left[\varphi\left(LR\left(\eta\vert m,x;S,\theta\right)\right)\right]\geq{\rm E}_{m,x}\left[\varphi\left(LR\left(\eta\vert m,x;NS,\theta\right)\right)\right].
\]
This comes from the fact that a sourced tweet is just more likely
to be right, so the likelihood ratio will be higher in distribution
so for every originator and type of $F$, sourced tweets have more
value in distribution so $Q_{S}\succ Q_{NS}$. 

Again, if we assume sources are fully revealing $\alpha_{S}\rightarrow1$
but without a source we have $\underline{v}^{NS}<0$. Retweeting shares
are given by $\widehat{G}_{S}\left(0\right)$ and $\frac{1}{2}G_{NS,H}\left(\underline{v}^{NS}\right)+\frac{1}{2}G_{NS,L}\left(\underline{v}^{NS}\right)$
under sourcing and no sourcing, respectively.

Crucially, even assuming sources are intrinsically good, retweeting
can be reduced. This comes from the fact that the perception payoff
effect can simply outweigh the gains in quality. If there is a source
there is nothing to signal, whereas if there is no source $F$ has
a signaling motivation that is traded off against quality.
\begin{rem}
	A natural question to ask is whether the since the arguments for celebrity
	versus Joe/Jane and sourced versus unsourced are identical, if anything
	seemingly relabeling, then the effects of sourced messaging and celebrity
	origination must have the same sign. But more careful reflection demonstrates
	that this is not true. Recall that retweeting share can be non-monotonic
	in $\alpha$ in this model. That is, given an initial $\alpha$, a
	move to some $\alpha'>\alpha$ can lead to a decline in retweeting
	share and whether this is the case can depend on $\left(\pi_{H},\pi_{L},\lambda\right)$.
	Concretely, recall the case of $\left(\pi_{H}=0.65,\pi_{L}=0.5,\lambda=50\right)$
	in Figure \ref{fig:retweet-sim} where the retweet share is non-monotonic
	with $\alpha$. Thus, the increase due to a celebrity versus the increase
	due to adding a source need not be the same and in fact can generate
	different signs on retweeting behavior. 
\end{rem}

\subsubsection{Myths}

We discuss the case of myths briefly as the argument is similar, though
not identical, to the above argument. A myth can be thought of as
a topic where the priors (i.e., the informativeness of the private
signals in the language above) are worse for all types: $\pi_{\theta}^{myth}<\pi_{\theta}^{none}$.
Of course, it may be the case that more able types face a relatively
smaller decline in prior belief quality relative to less able types,
or it may be the opposite. Following the same arguments as above,
just as the model can be non-monotonic in $\alpha$, moving to myths
from non-myth topics can increase or decrease the retweet share. To
see the intuition, if myths ensure that neither type has strong priors
to begin with, then there is no signaling impetus whatsoever. So the
quality of their signals and the quality of the message entirely drives
the retweet share. In contrast, if there is considerable scope for
signaling as high types are differentially more able to distinguish
myths (rather than non-myths) relative to low types, then retweeting
can increase.

\subsubsection{Exposure}

Finally, we look at the effect of having multiple historical exposures
to a topic.

The first simple way to think about this is as follows. Consider the
case where $F$ has received (and for the sake of discussion passed
on) multiple messages $m_{t}$ for $m_{1},...,m_{k-1}$ about $\eta$
in the past. Then the prior coming into the $k$th event is closer
to the truth and the marginal value of $m_{k}$ is lower, thereby
reducing the instrumental value to $F$ to passing on $m_{k}$. This
can make the retweeting share concave in $k$.

There is a second way to think about this, further incorporating the
signaling component of the model. To see this, suppose that every
message is about independent concepts (so $\eta_{t}$ is independent
every period $t$), so the commonality between all prior exposures
is mediated only through $F$'s followers' perception of $F$'s types.
That is, every instance of retweeting a message updates $F$'s followers
beliefs about $F$'s type which then serves as the prior going into
the next period. Exposure then affects the beliefs in this dynamic
manner.

As above let $k$ denote the number of prior exposures and $\Pr_{k}\left(a=H\right)$
be the prior in round $k+1$. We can see that $\Pr_{k}\left(a=H\right)\rightarrow1_{\left\{ a=H\right\} }$
as $k\rightarrow\infty$; that as the number of rounds tend to infinity,
the type is revealed even in the signaling game. This can be seen
as a consequence of the martingale convergence theorem. In this way,
signaling is self-limiting. Loosely speaking, what this means is that
with many exposures the signaling impetus can be small (which is equivalent
to the $\lambda=0$ case). In this case individuals only retweet positive
information, so relative to cases with a signaling impetus, it is
even possible for the retweet rate to decline.

All told, the effect of exposures therefore is an empirical matter
in terms of whether and when the marginal exposure does not matter
and even whether it can adversely affect retweeting rates.

\clearpage

\clearpage
\section{Does content affect retweeting?}\label{sec:rt_content}

\setcounter{table}{0}
\renewcommand{\thetable}{B.\arabic{table}}
\setcounter{figure}{0}
\renewcommand{\thefigure}{B.\arabic{figure}}

\begin{table}[!h]
	\centering
	\caption{How Content Affects Retweeting by $F_1$ likes/retweets}\label{tab:content_F1}
	\scalebox{0.9}{\begin{threeparttable}
			\begin{tabular}{lccccc} \hline
 & (1) & (2) & (3) & (4) & (5) \\
 & Poisson & Poisson & Poisson & Poisson & Poisson \\
VARIABLES & \# Pooled & \# Retweets & \# Likes & \# Pooled & \# Retweets \\ \hline
 &  &  &  &  &  \\
 Myth-busting Facts & 0.588 & 0.627 & 0.518 & -0.0481 & 0.136 \\
 & (0.319) & (0.346) & (0.381) & (0.296) & (0.413) \\
 & [0.0654] & [0.0698] & [0.174] & [0.871] & [0.742] \\
Access Info & 0.402 & 0.319 & 0.530 & 0.315 & 0.477 \\
 & (0.258) & (0.292) & (0.309) & (0.294) & (0.366) \\
 & [0.118] & [0.275] & [0.0863] & [0.284] & [0.192] \\
Importance Info & 0.543 & 0.526 & 0.565 & 0.466 & 0.442 \\
 & (0.229) & (0.267) & (0.290) & (0.246) & (0.343) \\
 & [0.0178] & [0.0487] & [0.0516] & [0.0578] & [0.197] \\
Celeb writes and tweets &  &  &  & 1.040 & 1.327 \\
 &  &  &  & (0.283) & (0.374) \\
 &  &  &  & [0.000242] & [0.000382] \\
 Myth $\times$ Celeb Direct &  &  &  & 0.652 & 0.432 \\
 &  &  &  & (0.381) & (0.485) \\
 &  &  &  & [0.0871] & [0.374] \\
Access $\times$ Celeb Direct &  &  &  & -0.0101 & -0.299 \\
 &  &  &  & (0.374) & (0.451) \\
 &  &  &  & [0.979] & [0.508] \\
Importance $\times$ Celeb Direct &  &  &  & 0.0558 & 0.00945 \\
 &  &  &  & (0.314) & (0.406) \\
 &  &  &  & [0.859] & [0.981] \\
 Myth $\times$ Celeb RT Org &  &  &  & 0.250 & 0.147 \\
 &  &  &  & (0.290) & (0.384) \\
 &  &  &  & [0.389] & [0.701] \\
Access $\times$ Celeb RT Org &  &  &  & 0.103 & -0.0418 \\
 &  &  &  & (0.244) & (0.256) \\
 &  &  &  & [0.672] & [0.870] \\
Importance $\times$ Celeb RT Org &  &  &  & 0.135 & 0.402 \\
 &  &  &  & (0.215) & (0.190) \\
 &  &  &  & [0.531] & [0.0348] \\

 &  &  &  &  &  \\
Observations & 492 & 492 & 492 & 492 & 492 \\
 Depvar Mean & 3.644 & 3.644 & 3.644 & 3.644 & 3.644 \\ \hline
\end{tabular}

			\begin{tablenotes}
				Notes: Standard errors (clustered at the celebrity/organization level) are reported in parentheses. $p$-values are reported in brackets. All columns include fixed effects for number of non-exception tweets assigned and condition on non-exception tweets. The omitted category is non-myth facts.
			\end{tablenotes}
	\end{threeparttable}}
\end{table}

\clearpage

\section{Does RT count affect retweeting?}\label{sec:rt_count}

\setcounter{table}{0}
\renewcommand{\thetable}{C.\arabic{table}}
\setcounter{figure}{0}
\renewcommand{\thefigure}{C.\arabic{figure}}

\begin{table}[!h]
	\centering
	\caption{Impact of No. of Forced Joe RTs on $F_2$ and $F_1$ likes/retweets}\label{tab:exjoegroup_F1_F2}
	\scalebox{1}{\begin{threeparttable}
			\begin{tabular}{lcc} \hline
 & (1) & (2) \\
 & F2 & F1 \\
 & Poisson & Poisson \\
VARIABLES & \# Retweets & \# Retweets \\ \hline
 &  &  \\
5 Forced Joe RTs assigned & 0.0399 & 0.444 \\
 & (0.346) & (0.388) \\
 & [0.908] & [0.252] \\
10 Forced Joe RTs assigned & 0.244 & 0.0395 \\
 & (0.414) & (0.440) \\
 & [0.556] & [0.928] \\
15 Forced Joe RTs assigned & 0.256 & 0.207 \\
 & (0.407) & (0.359) \\
 & [0.529] & [0.565] \\
 &  &  \\
Observations & 505 & 184 \\
Phase Control & \checkmark & \checkmark \\
Log \#followers control & \checkmark & \checkmark \\
Message style control & \checkmark & \checkmark \\
Depvar Mean & 0.184 & 2.707 \\
 1 Forced Joe RT assigned log mean & -2.331 & 0.870 \\ \hline
\end{tabular}

			\begin{tablenotes}
				Notes: Robust standard errors are reported in parentheses. $p$-values are reported in brackets.
			\end{tablenotes}
	\end{threeparttable}}
\end{table}

\clearpage

\section{Effect of celebrity retweeting organizations}\label{sec:orgs}

\setcounter{table}{0}
\renewcommand{\thetable}{D.\arabic{table}}
\setcounter{figure}{0}
\renewcommand{\thefigure}{D.\arabic{figure}}

\begin{table}[!h]
	\centering
	\caption{Reach vs. Endorsement: Value of Celeb Endorsement for Joes and Organizations through Involvement measured by $F_2$ likes/retweets}\label{tab:celeb_v_org_v_joe_F2_appendix}
	\scalebox{0.85}{\begin{threeparttable}
			\begin{tabular}{lccc} \hline
 & (1) & (2) & (3) \\
 & Poisson & Poisson & Poisson \\
VARIABLES & \# Pooled & \# Retweets & \# Likes \\ \hline
 &  &  &  \\
Celeb writes and tweets & 0.423 & 0.421 & 0.574 \\
 & (0.182) & (0.179) & (0.520) \\
 & [0.0201] & [0.0185] & [0.269] \\
Org writes and Celeb retweets & 0.564 & 0.600 & 0.255 \\
 & (0.221) & (0.258) & (0.520) \\
 & [0.0107] & [0.0200] & [0.624] \\
 &  &  &  \\
Observations & 1,791 & 1,791 & 1,791 \\
 Joe writes mean & 0.0417 & 0.0343 & 0.00745 \\ \hline
\end{tabular}

			\begin{tablenotes}
				Notes: Standard errors (clustered at the original tweet level) are reported in parentheses. $p$-values are reported in brackets. The sample conditions on tweets that are not sensitive and includes tweets originated by Joes, organizations, and celebrities.  All regressions control for phase, celebrity fixed effects, and content fixed effects.
			\end{tablenotes}
	\end{threeparttable}}
\end{table}

\begin{table}[!h]
	\centering
	\caption{Value of Celeb Endorsement for Joes and Organizations through Composition measured by $F_1$ likes/retweets}\label{tab:celeb_v_org_v_joe_F1_appendix}
	\scalebox{0.85}{\begin{threeparttable}
			\begin{tabular}{lccc} \hline
 & (1) & (2) & (3) \\
 & Poisson & Poisson & Poisson \\
VARIABLES & \# Pooled & \# Retweets & \# Likes \\ \hline
 &  &  &  \\
Celeb writes and tweets & 1.205 & 1.411 & 0.913 \\
 & (0.102) & (0.105) & (0.131) \\
 & [0] & [0] & [0] \\
Org writes and Celeb retweets & 0.0597 & 0.246 & -0.208 \\
 & (0.135) & (0.128) & (0.180) \\
 & [0.658] & [0.0540] & [0.247] \\
 &  &  &  \\
Observations & 452 & 452 & 452 \\
 Joe writes and Celeb retweets mean & 2.058 & 1.045 & 1.013 \\ \hline
\end{tabular}

			\begin{tablenotes}
				Notes: Standard errors (clustered at the original tweet level) are reported in parentheses. $p$-values are reported in brackets. The sample conditions on tweets that are not sensitive and includes tweets originated by Joes, organizations, and celebrities.  All regressions control for phase, celebrity fixed effects, and content fixed effects.
			\end{tablenotes}
	\end{threeparttable}}
\end{table}

\clearpage

\section{Do opinions change?}\label{sec:opinions}

\setcounter{table}{0}
\renewcommand{\thetable}{E.\arabic{table}}
\setcounter{figure}{0}
\renewcommand{\thefigure}{E.\arabic{figure}}

\begin{table}[!h]
	\caption{Was there opinion change?}\label{tab:opinion_immun}
	\centering
	\scalebox{1}{\begin{threeparttable}

			\begin{tabular}{lcccc} \hline
 & (1) & (2) & (3) & (4) \\
 & OLS & Ologit & Ologit & Ologit \\
VARIABLES & Opinion Index & Beneficial & Safe & Importance \\ \hline
 &  &  &  &  \\
Std. Exposure to tweets & 0.00900 & -0.0795 & 0.0758 & 0.0430 \\
 & (0.0292) & (0.0703) & (0.0637) & (0.0753) \\
 & [0.758] & [0.259] & [0.234] & [0.568] \\
 & \{.888\} & \{.469\} & \{.611\} & \{.642\} \\
 &  &  &  &  \\
Observations & 2,441 & 2,441 & 2,441 & 2,441 \\
Potential exposure control & \checkmark & \checkmark & \checkmark & \checkmark \\
Double Post-LASSO & \checkmark & \checkmark & \checkmark & \checkmark \\
 Depvar Mean & 0.00 & 3.679 & 3.409 & 3.752 \\ \hline
\end{tabular}

			\begin{tablenotes}
				Notes: Standard errors (clustered at the combination of celebs followed level) are reported in parentheses. Clustered $p$-values are reported in brackets. Randomization inference (RI) $p$-values are reported in braces. Demographic controls include age, sex, province, dummy for urban area and dummy for having children. One standard deviation of exposure is 14.96 tweets.  Column 1 presents a (standardized) index from the first principal component of a PCA decomposition of the three indices.
			\end{tablenotes}
	\end{threeparttable}}
\end{table}

\end{document}